\documentclass[aps,prb,twocolumn,superscriptaddres]{revtex4-1}
\usepackage{times}
\usepackage{graphicx}
\usepackage{dcolumn}
\usepackage{bm}
\usepackage{color}
\usepackage{amsmath}
\usepackage{amssymb}
\usepackage{braket}
\newcommand\cdks[3]{c^{#1}_{#2, #3\sigma}}
\newcommand\fdks[3]{f^{#1}_{#2, #3\sigma}}
\newsavebox{\mybox}

\usepackage[colorlinks=true]{hyperref}

\begin{document}

\title{Pole structure of the electronic self-energy with coexistence of Charge order and Superconductivity}

\author{M. Grandadam}

\affiliation{Institut de Physique Th\'eorique, Universit\'e Paris-Saclay, CEA, CNRS, F-91191 Gif-sur-Yvette, France}

\author{C. P\'epin}

\affiliation{Institut de Physique Th\'eorique, Universit\'e Paris-Saclay, CEA, CNRS, F-91191 Gif-sur-Yvette, France}

\begin{abstract}
We compare the pole structure of the electronic Green's function obtained by Cluster Dynamical Mean Field Theory to the results from the fractionalized Pair Density Wave idea. In the superconducting phase, we can consider the system in a state with coexistence of Superconducting and Charge order. Writing the Green's function in a way analogous to the previously proposed ``hidden-fermions'' model from Ref.[\onlinecite{Sakai16}] leads to a similar pole structure for the self-energy. The fractionalization of the Pair Density Wave order also describes the pseudogap phase as a superposition of superconducting and charge order fluctuations. Considering a phenomenological lifetime for the particle-particle and particle-hole pairs leads to an electronic spectral function that matches the numerical results.
\end{abstract}

\maketitle

\section{Introduction}\label{sec:Intro}
Since their discovery, cuprate superconductors have been the focus of many investigations to understand the nature of the relation between the high-temperature superconducting phase and the pseudogap that appears when they are doped away from half-filling\cite{Alloul89,Warren89}. They have been the subject of an important amount of work on the theoretical, experimental and numerical side. In the latter case, a lot of efforts have been invested into the development of the Dynamical Mean Field Theory that allows for solving the Hubbard model by mapping it onto an impurity problem\cite{Georges:1996tm}. The technique has been extended to include multiple sites in the Cluster Dynamical Mean Field Theory (CDMFT)\cite{Kotliar01,Maier05} and gives precise results on the energy dependence of the electronic Green's function at specific points of the Brillouin zone\cite{Gull10,Sordi:2010iw,Sordi:2012bb}. Recent results from CDMFT on 2x2 clusters have shown a peculiar link between the structure of the electronic self-energy in the pseudogap (PG) and in the superconducting (SC) phase\cite{Sakai16,Sakai:2016bo,Sakai:2018jq} that deviate from the standard BCS theory. Multiple propositions have been made to explain this with an analytical model such as a hidden-fermion model\cite{Sakai16} or a $SU\left(2\right)$ gauge theory\cite{Scheurer18a}.\\
On the experimental side, there is growing evidence that the charge ordering observed above the superconducting temperature\cite{Doiron-Leyraud07,Sebastian12,Chang12,Blanco-Canosa13,Blackburn13a,Ghiringhelli12,Gerber:2015gx,Chang:2016gz,Wu11,Wu:2015bt,Julien15} plays an important role in the pseudogap physics. In fact, recent Raman experiment\cite{Loret19} suggests that the Charge Density Wave (CDW) gap is of the same order as the superconducting gap. This charge order is also known to coexist with the superconductivity at low temperature and long-range CDW is observed when superconductivity is destroyed by a magnetic field\cite{Gerber:2015gx,Chang:2016gz,LeBoeuf13}. A recent proposal linked these two orders by the fractionalization of modulated particle-particle pairs, \emph{i.e.} of a Pair Density Wave (PDW) order\cite{Chakraborty19,Grandadam20a}, leading to a constraint between the resulting CDW and SC amplitudes. The resulting superposition of fluctuations in both channels was used to explain\cite{Grandadam20b} the momentum, energy and temperature dependence of the electronic spectral function observed in Bi2201 by angle-resolved photo-emission spectroscopy\cite{He11} (ARPES).\\
We are here interested in the possibility that the fine structure in the self-energy obtained by CDMFT calculation\cite{Haule07,Sakai16,Gull:2013hh} on the Hubbard model could be explained by the same scenario. In the superconducting phase, the electronic Green's function is given by considering a coexistence of SC and CDW order. We write the electronic Green's function in the superconducting phase in a way analogous to the hidden-fermion proposition that was used to describe CDMFT results\cite{Sakai16,Sakai:2016bo,Sakai:2018jq} by coupling electrons to ``hidden" fermionic excitations on top of pairing. We identify the hidden-fermion with a modulating order and show that it leads to a pole structure in the self-energy close to the one obtained via numerical calculation below $T_c$. The fractionalized PDW scenario gives a description of the pseudogap that relies on the superposition of SC and CDW fluctuations in the antinodal region and is different from the hidden-fermion model mentioned previously. We show that the fractionalized PDW still reproduces the electronic Green's function in the pseudogap phase.
\\
\section{Electronic spectral function from CDMFT}
Cluster extension of CDMFT with 2x2 sites gives detailed energy-dependence of the electronic Green's function in the Hubbard model at low doping. Due to the small system size, this is however limited to some specific points of the Brillouin zone. One of the most important results concerns the anti-nodal point at $k=\left(0,\pi\right)$ where the pseudogap is observed above the superconducting state. CDMFT allows us to decompose the deviation from the non-interacting Green's function into two parts : the normal and anomalous self-energy. The final form of the electronic Green's function is given by :
\begin{align}
G \left(k,\omega \right) &= \left[\omega-\epsilon_k-\Sigma_N \left(k,\omega \right)-W \left(k,\omega \right)\right]^{-1}, \nonumber \\
W\left(k,\omega\right) &= \frac{\Sigma_{AN} \left(k,\omega \right)^2}{\omega+\epsilon_k-\Sigma_N \left(k,-\omega \right)^*} \label{green}
\end{align}
The anomalous part of the self-energy is directly related to the superconducting order and vanishes outside of the ordered phase but the normal part presents features that lead to the pseudogap in the anti-nodal region at higher temperature. Another quantity of interest is the electronic spectral function $A\left(k,\omega\right) = \frac{-1}{\pi}\text{Im}\left(G \left(k,\omega \right)\right)$ which relate directly to ARPES experiment for occupied states ($\omega<0$). Interestingly there are strong links between the structure in the normal self-energy above and below $T_c$ as well as connection with the anomalous self-energy. These seems to indicate that the pseudogap physics is strongly related to the superconducting state.\\

\begin{figure}
\includegraphics[width = 9 cm]{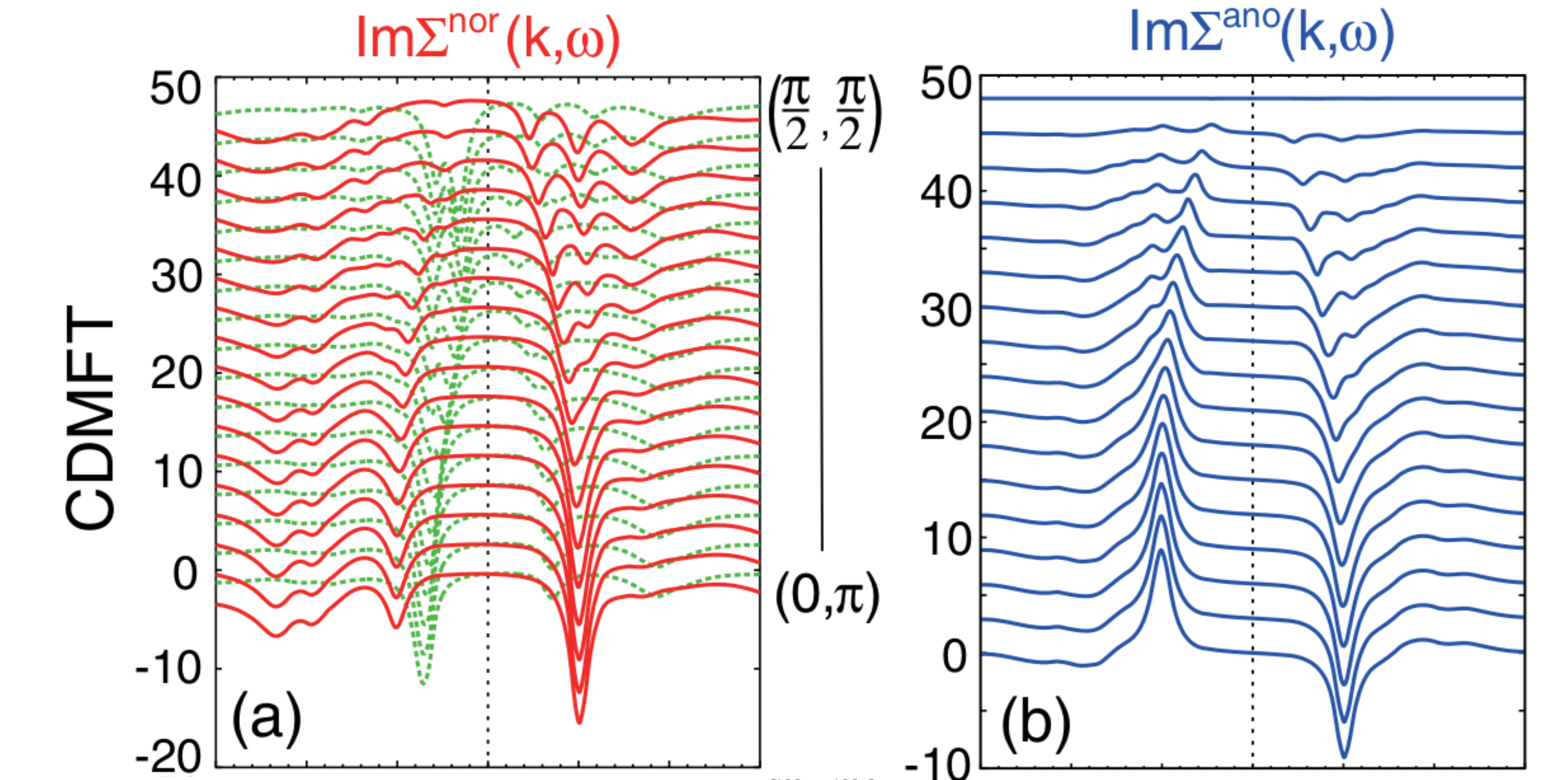}
\caption{\textbf{(a)} Normal self-energy (red line) and electronic spectral function (green dotted lines) obtained by CDMFT for momenta on the line $k = \left(0,\pi\right) - \left(\pi/2,\pi/2\right)$. The self-energy shows two isolated poles in the anti-nodal region at $\omega = \pm \omega_0$ but with a clear asymmetry in weight. In the nodal region the peak at positive energy splits. The pole at negative energy is damped is difficult to follow.\textbf{(b)} Anomalous self-energy obtained by CDMFT for the same momenta. It shows pairs of poles at the same position as in the normal part but with an anti-symmetric weight. The same splitting occurs when going from the anti-nodal to the nodal region. The weight close to $k=\left(\pi/2,\pi/2\right)$ is close to zero, indicating that the coupling to the hidden-fermion is specific to the anti-nodal region.}
\label{fig:fig1}
\end{figure}

The energy dependence of the normal self-energy (red lines) and of the electronic spectral function (green dotted line) for momenta on a line from $k=\left(0,\pi\right)$ to $k=\left(\pi/2,\pi/2\right)$ is shown in Fig.\ref{fig:fig1}(a) in the superconducting phase.The distinctive feature of the self-energies in the anti-nodal region (at $k=\left(0,\pi\right)$) is the two symmetric peaks at low energy. The two poles of $\Sigma_N$ have a marked weight asymmetry with the pole at $\omega > 0$ having a larger spectral weight than the one at negative energies. The spectral function plotted at the same momentum (green dotted line in Fig.\ref{fig:fig1}(a)) presents 3 poles, one at negative energy which is well defined and 2 at positive energies which are broader. In contrast, the two poles of $\Sigma_{AN}$ shown in Fig.\ref{fig:fig1}(b) for the same momenta have the same position as the poles of $\Sigma_N$ but opposite weight. When going toward the nodal region we observe that the pole at positive energy split into two poles of approximately equal weight. This is also the case for the pole at negative energy even though the vanishing weight close to $k=\left(\pi/2,\pi/2\right)$ makes it harder to pinpoint the position of the two peaks.\\
Another crucial information given by CDMFT is the cancellation that occurs between the poles of $\Sigma_N$ and $W$. Indeed it has been observed that, besides having poles at the same energy, the residue of both the normal and the anomalous the self-energy at these poles is such that they cancel in the expression of the single-particle Green's function. This has consequences for the electronic spectral function as the poles of the self-energy should correspond to zeros of the spectral function, effectively leading to the splitting of the non-interacting band.\\
These results were interpreted in terms of "hidden-fermions" which couple to the original electronic degree of freedom and are themself susceptible to pairing\cite{Sakai16,Sakai:2016bo,Sakai:2018jq}. It is then possible to write a mean-field Hamiltonian for this model and extracts analytical expression for the normal and anomalous self-energy. As we will show in the next section, this construction guarantees the cancellation between $\Sigma_N$ and $W$ mentioned earlier. It also displays a structure for both parts of the self-energy in agreement with the numerical results with proper choice for the hidden-fermion. We will present the model and its main features in the next section and then show how to incorporate charge order in the hidden-fermion formalism and compare the results to CDMFT in both the superconducting and the pseudogap state.

\section{Hidden-fermion Model}
We start with the same hidden-fermion model from Ref.\onlinecite{Sakai16}. This model describe electrons ($\cdks{\dagger}{k}{}$) on a square lattice coupled to other fermionic excitation ($\fdks{\alpha \dagger}{k}{}$) :
\begin{align}
H = &\sum_{k,\sigma} \epsilon_k \cdks{\dagger}{k}{}\cdks{}{k}{} + \sum_{k,\sigma} \left(\sigma\Delta_k \cdks{\dagger}{k}{}\cdks{\dagger}{-k}{-} + h.c\right) \nonumber \\
& + \sum_{\alpha,k,\sigma} \epsilon^{f,\alpha}_k \fdks{\alpha \dagger}{k}{}\fdks{\alpha}{k}{} + \sum_{\alpha,k,\sigma} \left(\sigma\Delta^{f,\alpha}_k \fdks{\alpha \dagger}{k}{}\fdks{\alpha \dagger}{-k}{-} + h.c\right) \nonumber \\
& + \sum_{\alpha,k,\sigma} \left(V^{\alpha}_k \cdks{\dagger}{k}{}\fdks{\alpha}{k}{}+h.c\right) \label{eq:H}
\end{align}
The electronic spectral function can then be obtain and the coupling to the different $\fdks{\alpha}{k}{}$ excitation leads to a normal and anomalous self-energy given by :
\begin{align}
&\Sigma_{N}\left(k,\omega\right) = \sum_{\alpha} \frac{{V_k^{\alpha}}^2 \left(\omega+\epsilon_k^{f,\alpha}\right)}{\omega^2-{\epsilon_k^{f,\alpha}}^2-{\Delta_k^{f,\alpha}}^2} \nonumber \\
&\Sigma_{AN}\left(k,\omega\right) = \Delta_k + \sum_{\alpha} \frac{{-V_k^{\alpha}}^2\Delta_k^{f,\alpha}}{\omega^2-{\epsilon_k^{f,\alpha}}^2-{\Delta_k^{f,\alpha}}^2} \label{eq:self}
\end{align}
As can be seen from Eq.\eqref{eq:self} the normal and anomalous part of the self-energy share a similar pole structure which is primarily governed by the choice of $\Delta^{\alpha}_k$ and $\epsilon^{\alpha}_k$. The value of the coupling between the $\cdks{\dagger}{k}{}$ and the $\fdks{\alpha \dagger}{k}{}$ fermions only enters in the spectral weight at the different poles as can be seen by computing the residue of the self-energy :
\begin{align}
&\text{Res}\left(\Sigma_N,\pm \omega^{\alpha}\right) = \frac{{V_k^{\alpha}}^2}{2}\left(1 \pm \frac{\epsilon_k^{f,\alpha}}{\sqrt{{\epsilon_k^{f,\alpha}}^2+{\Delta_k^{f,\alpha}}^2}}\right) \nonumber \\
&\text{Res}\left(\Sigma_{AN},\pm \omega^{\alpha}\right) = \mp \frac{{V_k^{\alpha}}^2\Delta_k^{\alpha}}{2 \omega^{\alpha}} \label{eq:residue}
\end{align}
We can see that the self-energies have poles due to the hidden-fermions at the energies $\omega^{\alpha} = \pm \sqrt{{\epsilon^{f,\alpha}_k}^2+{\Delta_k^{f,\alpha}}^2}$. There is however an important distinction between the normal part and the anomalous part of the self-energy in the weight associated with each of these poles. In fact, the anomalous self-energy has equal and opposite weight at each pole while it depends on the value of $\epsilon^{f,\alpha}_k$ for the normal self-energy. In particular the sign of $\epsilon^{f,\alpha}_k$ will determine which of $\omega^{\alpha}$ or $-\omega^{\alpha}$ has a higher weight. It is also possible to check the cancellation mentioned previously :
\begin{align}
\text{Res}\left(W,\pm \omega^{\alpha}\right) &= \text{Res}\left(\frac{\Sigma_{AN} \left(k,\omega \right)^2}{\omega+\epsilon_k-\Sigma_N \left(k,-\omega \right)^*},\pm \omega^{\alpha}\right) \nonumber \\
&= -\frac{{V_k^{\alpha}}^2}{2}\left(1 \pm \frac{\epsilon_k^{f,\alpha}}{\sqrt{{\epsilon_k^{f,\alpha}}^2+{\Delta_k^{f,\alpha}}^2}}\right) \nonumber \\
&= -\text{Res}\left(\Sigma_N,\pm \omega^{\alpha}\right) \label{eq:cancel}
\end{align}
\\
\indent The original proposition was that only one of these hidden fermionic excitations is relevant to explain the structure of the self-energy at low energy and that identifying $\epsilon^f_k = z_k\epsilon_{k+\bm{\pi}}-\mu_f$, where $z_k$ is the renormalization factor due to the strong interactions and $\bm{\pi}=\left(\pi,\pi\right)$, leads to a pole structure analogous to the CDMFT results.
\begin{figure}
\includegraphics[width = 7 cm]{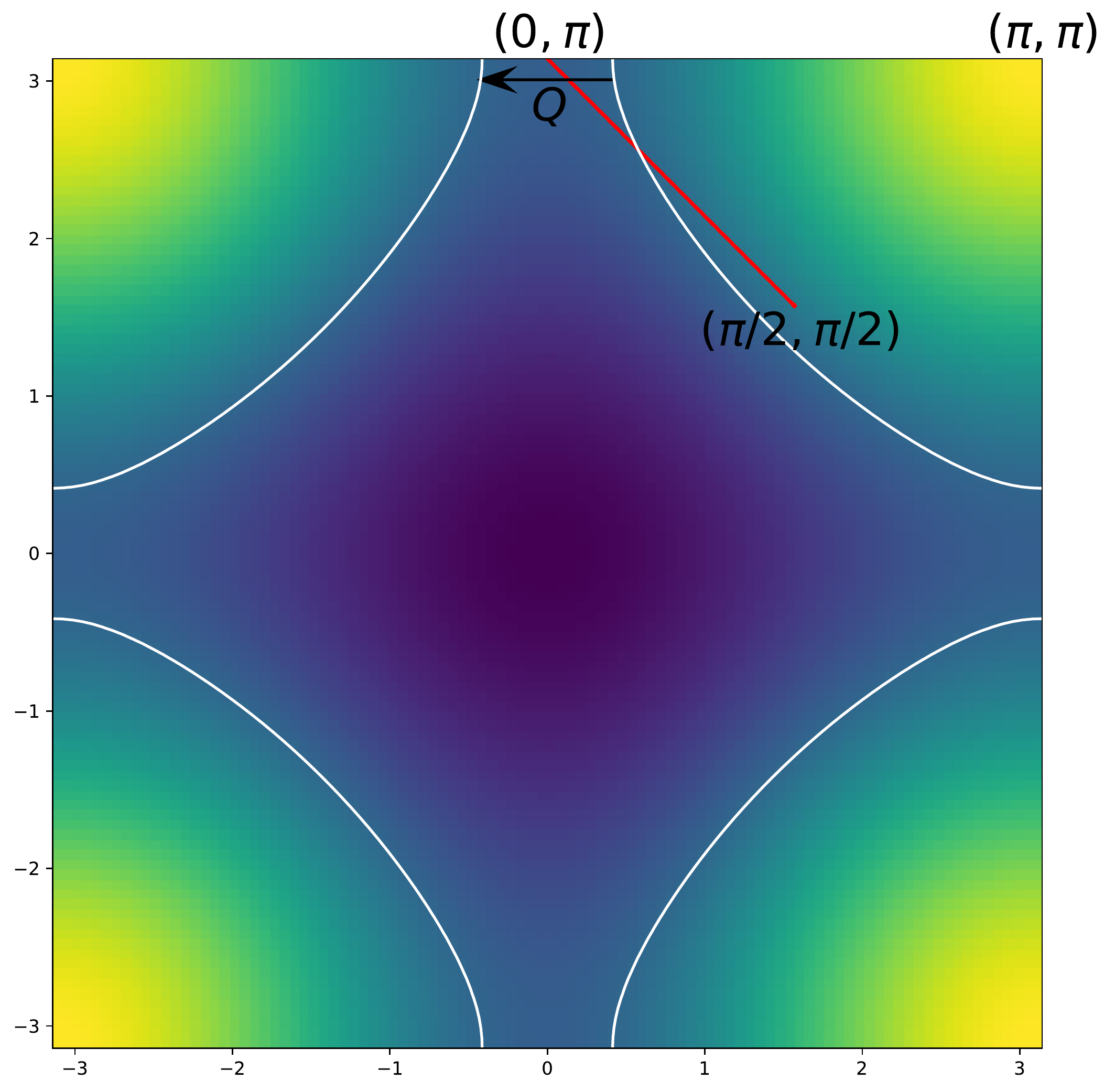}
\caption{Fermi surface for free electrons with dispersion given by $\epsilon_{k} = -2t \left( \cos\left(k_x\right)+\cos\left(k_y\right) \right) -4t^{\prime}\cos\left(k_x\right)\cos\left(k_y\right)-\mu$ where $t=1$, $t^{\prime}=-0.2$ and $\mu = -0.56$. The black arrow indicate the charge order wave-vector taken to link different part of the Fermi surface while the red line shows the path between $k=\left(0,\pi\right)$ and $k=\left(\pi/2,\pi/2\right)$ where the self energy is computed. }
\label{fig:fig2}
\end{figure}
We show here that another choice of hidden fermion can lead to satisfactory comparison with the results obtained by CDMFT but have a different interpretation. Our choice of hidden fermions is based on the experimental observation of charge order in cuprates (by X-Ray\cite{Chang12,Blanco-Canosa13,Campi15,Blackburn13a,Ghiringhelli12,Gerber:2015gx,Chang16} or NMR\cite{Wu11,Wu13a,Wu:2015bt} measurements) that persists in the superconducting phase. This coexistence of CDW and SC has been shown to successfully reproduce the single-particle electronic Green's function observed in Bi2201 by ARPES\cite{Grandadam20b} and the softening observed in the phonon spectrum at $T_c$\cite{Sarkar20}. Our analysis is thus based on the previous hidden-fermion model Eq.\eqref{eq:H} but we consider that two fermionic excitation will contribute and identify them with $\fdks{\alpha \dagger}{k}{} = \cdks{\dagger}{k+Q}{}$  and $\fdks{\beta \dagger}{k}{} = \cdks{\dagger}{k-Q}{}$.\\
\\
We focus on an axial wave-vector that links different parts of the Fermi surface, \emph{i.e.} $Q=\left(\pm Q_0,0\right)$ as shown in Fig.\ref{fig:fig2}. We also take into account strong correlations by adding a renormalization factor $z\left(k,\omega\right)$ that will modify the dispersion of the hidden fermion so that $\epsilon^{f,\alpha}_k = z \epsilon_{k+Q}$ and $\epsilon^{f,\beta}_k = z \epsilon_{k-Q}$ where we took $z\left(k,\omega\right)$ to be a constant. The bare dispersion is given by taking nearest and second neighbours hopping $\epsilon_{k} = -2t \left( \cos\left(k_x\right)+\cos\left(k_y\right) \right) -4t^{\prime}\cos\left(k_x\right)\cos\left(k_y\right)-\mu$. We took $t=1$ as our energy scale, $t^{\prime} = -0.2$ and $\mu$ is chosen to fix the electron density to $n=0.95$ using the Luttinger sum rule in the non-interacting case. Lastly, as we are interested in the superconducting phase, we take the SC gap with a d-wave form factor $\Delta_k = \frac{\Delta_0}{2}\left(\cos\left(k_x\right)-\cos\left(k_y\right)\right)$ and the other gaps are given by $\Delta^{f,\alpha}_k = \Delta_{k+Q}$ and $\Delta^{f,\beta}_k = \Delta_{k-Q}$.

\section{Results}
\subsection{Normal self-energy in the superconducting state}

\begin{figure}
\includegraphics[width = 8.4 cm]{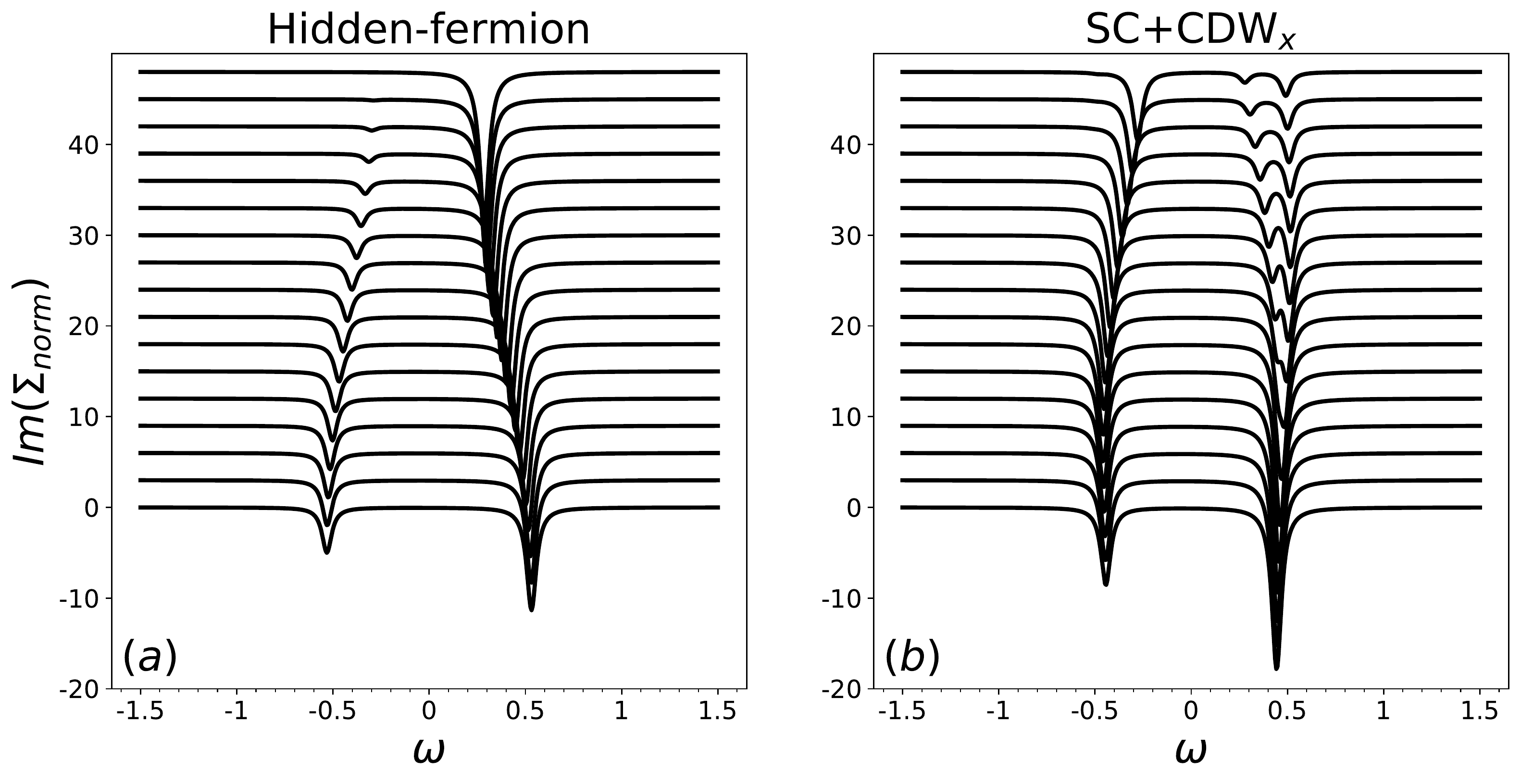}
\caption{\textbf{a} Normal self-energy in the hidden-fermion model of Ref.\onlinecite{Sakai16}. The asymmetry in the anti-nodal region is well reproduced. As there is only one hidden-fermion coupled to electrons it is not possible to recover the splitting in the nodal region. The weight of the positive energy pole is also maximum for $k=\left(\pi/2,\pi/2\right)$ in contrast to the numerical results shown in Fig.\ref{fig:fig2}. \textbf{(b)} Normal self-energy obtained while considering hidden-fermions due to CDW order with a wave-vector along the $x$ axis. Due to the symmetry between $\epsilon_{k+Q} = \epsilon_{k-Q}$ at $k=\left(0,\pi\right)$ we only have two visible poles with an asymmetry consistent with the CDMFT results. This symmetry is lost when going to the nodal region and we observe a splitting of the poles. The change of sign of $\epsilon_{k-Q}$ close to $k=\left(\pi/2,\pi/2\right)$ explains the significant weight for the pole at negative energy. We choose here $Q_0 = 0.27 \pi$, $z = 0.22$, $\Delta_0 = 0.55$, $V_0 = 0.7$ and a numerical broadening factor $i\eta = 0.03i$.}
\label{fig:fig3}
\end{figure}

To relate the different features of the normal self-energy to the properties of the hidden-fermions we need to study the structure of the normal part of the self-energy given by Eq.\eqref{eq:self}. The positions of the poles are given by $\omega^{\alpha}_{\pm} = \pm\sqrt{{\epsilon_k^{f,\alpha}}^2+{\Delta_k^{f,\alpha}}^2}$ and thus come by pair with symmetric position in energy. The weight associated with each pole is obtained by computing the residue at each of them and is given in Eq.\eqref{eq:residue}. The asymmetry in weight is thus controlled by the sign of $\epsilon^{f,\alpha}_k$ and the value of $\Delta^{f,\alpha}_k$. As noted in Ref.[\onlinecite{Sakai16}], the observation that the pole at positive energy has a larger weight than the pole at negative energy in the anti-nodal region impose the fact that $\epsilon^{f,\alpha}_{\left(0,\pi\right)}$ has to be positive. This constraint is naturally fulfilled in our model when we choose a CDW wave-vector larger than the anti-nodal Fermi wave-vector which also fits experimental observation. This is however no longer the case for the fermion identified with $\cdks{}{k-Q}{}$ when we approach the nodal region which explains the discrepancy between our model and the CDMFT results near $k=\left(\frac{\pi}{2},\frac{\pi}{2}\right)$.\\
The results for the original hidden-fermion idea\cite{Sakai16} that identifies $\fdks{\alpha \dagger}{k}{} = \cdks{\dagger}{k+\bm{\pi}}{}$ are shown in Fig.\ref{fig:fig3}(a) and match multiple characteristics of the numerical results. Notably, the asymmetry in the spectral weight in the anti-nodal region is well reproduced and the loss of weight for the peak at negative energy is found. Note that, following Eq.\eqref{eq:residue}, this happens when $\epsilon^{\alpha}_k$ is positive. There is thus a crucial role of the added $\mu_f$ term as the hidden-fermion dispersion $z_k \epsilon_{k+\pi}$ alone is negative for $k=\left(0,\pi\right)$. The combination of the renormalization induced by $z_k$ and the positive shift of $\mu_f$ gives an effective dispersion that is small relative to $\Delta_k$ and that does not change sign. However, because there is only one hidden fermion it is not possible to capture the splitting in the nodal region for $\omega>0$ and the weight of the positive peak is maximal at $k=\left(\pi/2,\pi/2\right)$ when the d-wave superconducting gap vanishes, in contrast to the numerical results.

We now turn to the fractionalized PDW hypothesis which gives the same form for the electronic Green's function, in the superconducting phase, to the hidden-fermion model with $\fdks{\alpha \dagger}{k}{} = \cdks{\dagger}{k+Q}{}$  and $\fdks{\beta \dagger}{k}{} = \cdks{\dagger}{k-Q}{}$. Our results for the two hidden fermions coming from the charge order are shown in Fig.\ref{fig:fig3}(b). At $k=\left(\pi,\pi\right)$ we can only see a pair of symmetric peaks due to the fact that $\epsilon_{k+Q} = \epsilon_{k-Q}$. Moreover, the charge order wave-vector being larger than the anti-nodal Fermi momentum means that we have $\epsilon_{k\pm Q}>0$. Using the result of Eq.\eqref{eq:residue}, this leads to the same weight asymmetry observed in the CDMFT study with the pole at $+\omega^{\alpha}$ having a higher weight than the pole at $-\omega^{\alpha}$. When going towards the nodal region the previous symmetry between $\epsilon_{k+Q}$ and $\epsilon_{k-Q}$ is lost and we observe a splitting of the poles at $\omega>0$ analogous to the numerical results. It is important to point out that the behaviours of the spectral weight for the two peaks are different. In fact, the pole due to the coupling to $\cdks{\dagger}{k+Q}{}$ will have a higher weight at $-\omega^{\alpha}$ in the nodal region. This is because $\epsilon_{k-Q}$ will change sign and become negative. Because the pairing gap for the hidden-fermion is taken as $\Delta_k^{f,\alpha}=\Delta_{k\pm Q}$, there is no cancellation of the weight at $k=\left(\pi/2,\pi/2\right)$ which result in a non-vanishing weight in the nodal region for the pole at $\omega<0$. Note that we took here a coupling between the electrons and the hidden-fermions to be independent of momentum, \emph{i.e} $V_k^{f,\alpha} = V_0$. This does not impact the position of the poles as shown by Eq.\eqref{eq:self} but only the weight at each pole (see Eq.\eqref{eq:residue}). Furthermore, results from CDMFT away from the antinodal point $k=\left(0,\pi\right)$ are extrapolated from the available momentum points. Thus, the discrepancy between the spectral weight in the nodal region can be due to multiple factors.\\\\

\subsection{Pseudogap phase}
We showed that both the antiferromagnetic hidden-fermion and the fractionalized PDW models give similar results in the superconducting phase. The two models differ strongly in the pseudogap phase where the hidden-fermion model restores the charge conservation symmetry but keeps the coupling between electrons and the hidden fermions unchanged. In contrast, the fractionalization of a Pair Density Wave leads to the superposition of SC and CDW fluctuations that have a strong impact on the electronic spectral functions.
\subsubsection{Pseudogap in the hidden-fermion model}
In the hidden-fermion model, the pseudogap phase is obtained by setting the superconducting order parameters $\Delta_k$ and $\Delta_k^{f,\alpha}$ to zero. The remaining part of the self-energy is due to the hybridization to the hidden-fermions through $V^{\alpha}$. The self-energy thus take the form:
\begin{align}
&\Sigma_{N}\left(k,\omega\right) = \sum_{\alpha} \frac{{V^{\alpha}_k}^2 }{\omega-{\epsilon^{f,\alpha}}_{k}} \nonumber \\
&\Sigma_{AN}\left(k,\omega\right) = 0 \label{eq:selfPG}
\end{align}

\begin{figure}
\includegraphics[width = 8 cm]{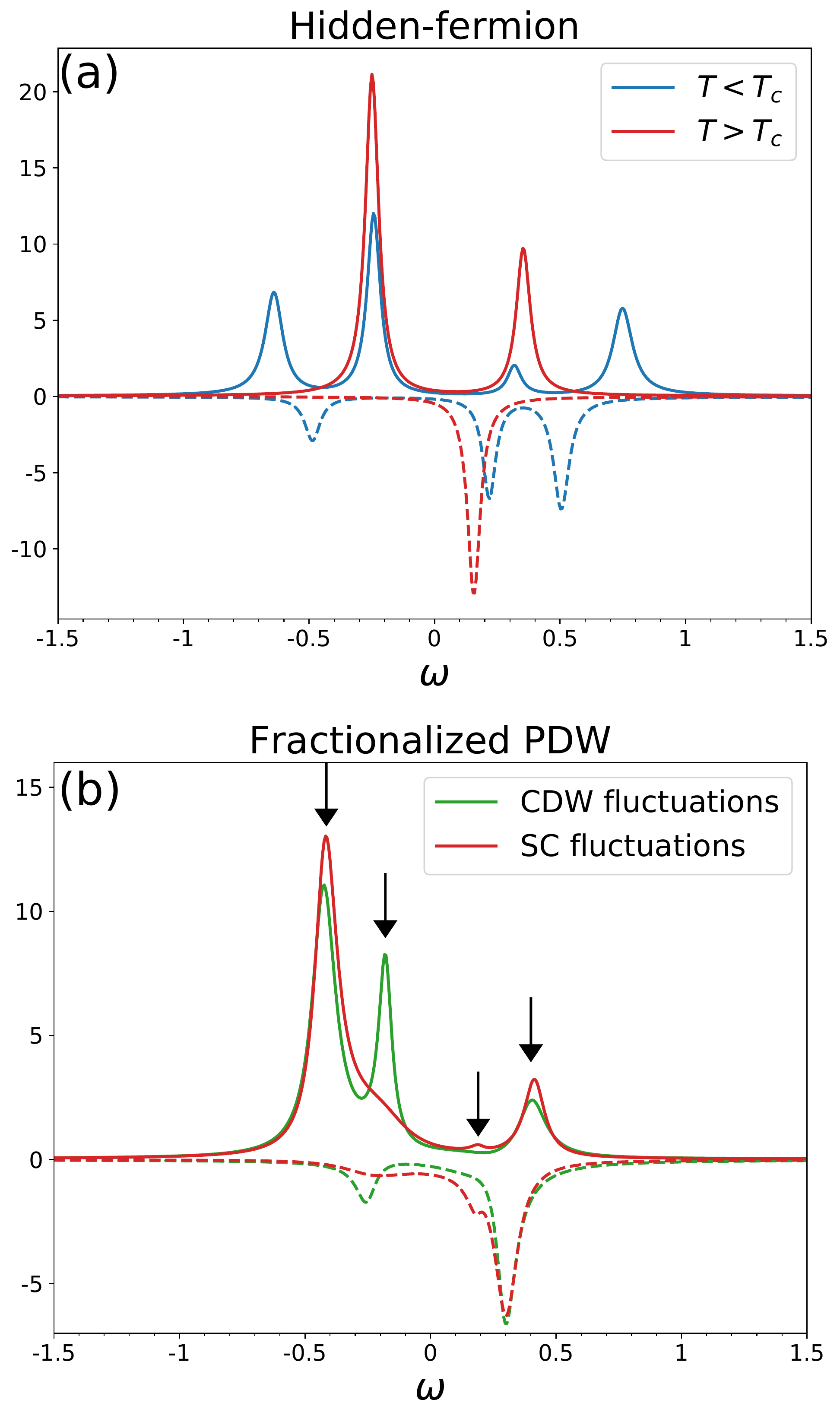}
\caption{\textbf{(a)} Electronic spectral function (full line) and self-energy (dotted line) in the pseudogap (red) and superconducting phase (blue) in the hidden-fermion model. Above $T_c$, the self-energy is reduced to a single pole at $\omega = \epsilon_{k\pm Q} > 0$ and the resulting electronic spectral function shows 2 poles. There is no spectral weight at $\omega=0$ but there will be gapless states away from $k_x=0$ in contrast to experimental observations. \textbf{(b)} Electronic spectral function (full line) and self-energy (dotted line) in the pseudogap phase from a fractionalized PDW at $k=\left(0,\pi\right)$. Green lines show effects of CDW fluctuations ($\Gamma_1$) while the red lines show the effect of superconducting fluctuations ($\Gamma_2$). Due to the superposition of SC and CDW in the fractionalized PDW idea, the electronic spectral function retains the 4 poles structure seen in the superconducting phase. Finite lifetime for the particle-particle or particle-hole pairs have a much stronger dampening effect in the two poles closest to $\omega=0$ which could explain the observed two peaks structure in CDMFT or the single occupied band in ARPES above $T_c$. We choose here $Q_0 = 0.27 \pi$, $z = 0.22$, $|\Psi_0| = V_0 = 0.15$, $\Gamma_1 = \Gamma_2 = 0.1$ and a numerical broadening factor $i\eta = 0.03i$.  }
\label{fig:fig4}
\end{figure}

The structure of the self-energy is then reduced to peaks located at the energy of the hidden-fermions as shown in Fig.\ref{fig:fig4}(a) (red dotted line). In our case, the self-energy in the anti-nodal region will display a single peak at $\omega > 0$ because $\epsilon_{\bm{\pi}+Q} = \epsilon_{\bm{\pi}-Q} > 0$. This leads to the electronic spectral function having two poles with a vanishing spectral weight at $\omega=0$ for $k=\left(0,\pi\right)$ (red line in Fig.\ref{fig:fig4}(a)). Note that a non-zero value of hybridization between $\cdks{}{k}{}$ and $\cdks{}{k+Q}{}$ should be interpreted as a long-range CDW at the mean-field level. As such, we should observe a Fermi surface reconstruction with gapless states in the pseudogap phase. This long-range order has only been observed by applicatying strong magnetic fields\cite{Gerber:2015gx,Chang:2016gz,LeBoeuf13} in the superconducting phase. Furthermore, the short-range charge order above $T_c$ is only observed below a transition temperature $T_{CO}$ which is lower than the pseudogap temperature $T^*$. The band structure observed by ARPES does not however show significant change for temperatures in the pseudogap region. It is thus inconsistent to consider that the pseudogap phase is solely due to the charge order.

\subsubsection{Fractionalized Pair Density Wave}
We start by giving details on the Pair Density Wave fractionalization as described in Ref.[\onlinecite{Chakraborty19,Grandadam20a}]. The fractionalization of the PDW operator is based on the relation between the $\eta$-mode, which is a modulated particle-particle pair and the SC ($\hat{\Delta}_{ij}=J\hat{d}_{ij}\sum_{\sigma}\sigma c_{i,\sigma}c_{j,-\sigma}$) and the CDW ($\hat{\chi}_{ij}=J\hat{d}_{ij}\sum_{\sigma}c_{i,\sigma}^{\dagger}c_{j,\sigma}e^{i\mathbf{Q.\left(\mathbf{r}_{i}+\mathbf{r}_{j}\right)/2}}$) operators where $J$ is the typical interaction strength giving rise to the SC and CDW orders. In particular we can write :
\begin{align}
\hat{\eta} & =[\hat{\Delta}_{ij},\hat{\chi}_{ij}^{\dagger}], &  & \hat{\eta}^{\dagger}=[\hat{\chi}_{ij},\hat{\Delta}_{ij}^{\dagger}],\label{eq:1}
\end{align}
where $\bm{Q}$ is the modulation wave-vector of the PDW. In a way analogous to the electron's fractionalization in strong coupling theory\cite{Baskaran88,Lee92}, the $\eta$-operators are then invariant with the following gauge structure
\begin{align}
\hat{\Delta}_{ij} & \rightarrow e^{i\theta}\hat{\Delta}_{ij}, &  & \hat{\chi}_{ij}\rightarrow e^{i\theta}\hat{\chi}_{ij}.\label{eq:2}
\end{align}
The effective field theory for the fluctuating PDW thus involved an emergent $U(1)$ gauge field whose fluctuation produces a constraint between the two fields:
\begin{align}
& |\hat{\Delta}_{ij}|^{2}+|\hat{\chi}_{ij}|^{2}\equiv\left|\Psi_{ij}\right|^{2}=\text{const},\label{eq:frac}
\end{align}
where $\Psi_{ij}=(\hat{\Delta}_{ij},\hat{\chi}_{ij})^t$.
The energy scale associated with Eq.(\ref{eq:frac}) is typically the scale at which the fractionalization occurs.\\

\begin{figure}
\includegraphics[width = 9 cm]{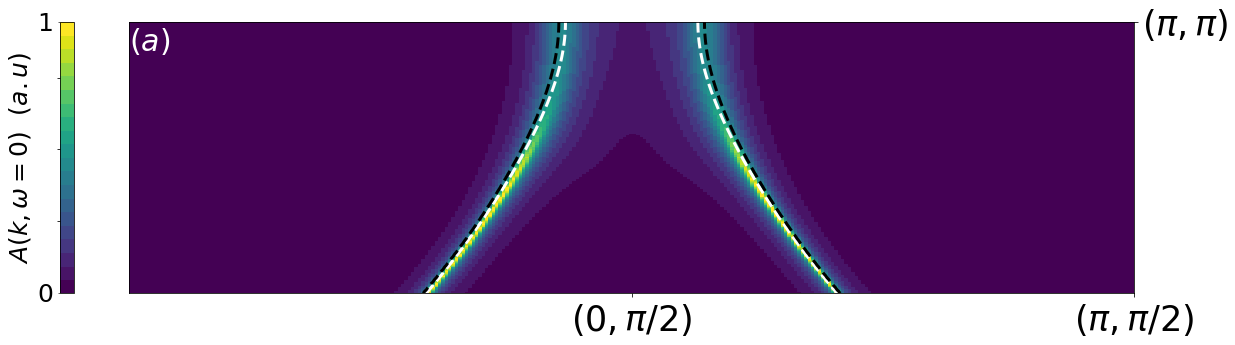}
\caption{\textbf{(a)} Electronic spectral function at $\omega = 0$ for the fractionalized PDW model. The white dotted line indicates the non-interacting Fermi surface while the black dotted line indicates the surface of zeros of the Green's function. We see that the Fermi surface is washed out in the antinodal region. We choose here $Q_0 = 0.27 \pi$, $z = 0.22$, $|\Psi_0| = 0.15$ and a numerical broadening factor $i\eta = 0.03i$.  }
\label{fig:fig5}
\end{figure}

Earlier descriptions of the pseudogap phase as a fractionalized PDW led to good agreement with experimental ARPES results for the electronic spectral function\cite{Grandadam20b}. The main characteristic of the pseudogap in this theory is to be an equal superposition of SC and CDW fluctuations in the anti-nodal regions. This is achieved at the mean-field level by considering the previous model but constraining the amplitude for the SC and CDW amplitude to be equal while adding finite lifetime $\Gamma_1$ and $\Gamma_2$ for the particle-hole and for the particle-particle pairs respectively. Because none of the parameters is put to zero, the self-energy and the spectral function have the same pole structure in both the pseudogap and the superconducting phases. In fact, the electronic spectral function at $k=\left(0,\pi\right)$ given in Fig.\ref{fig:fig4}(b) with both damping rate $\Gamma_1$ and $\Gamma_2$ being turned on alternatively clearly shows 4 poles indicated by arrows. The main observation is that there is a significant reduction in the spectral weight of the two poles at lower energy for both positive and negative energy. This is the same phenomenon that was used to describe the observation of a ``flat band'' below $T_c$ by ARPES in Bi2201\cite{Grandadam20b}. Similarly to this previous study, this description of the pseudogap effectively leads to Fermi arcs and no long-range order is expected. We show the resulting electronic spectral function at $\omega = 0$ in the antinodal region in Fig.\ref{fig:fig5}(b). We can see that part of the Fermi surface is washed out even in the absence of fluctuations ($\Gamma_1 = \Gamma_2 = 0$).\\
Despite having a similar structure for the self-energy in the pseudogap phase, there is an important difference between the structure of the Green's function in our approach and in previous studies based on fractionalization such as the one using a $SU(2)$ theory\cite{Scheurer18a} of fluctuating antiferromagnetism. In our case, the line of zeros for the real part of the Green's function (black dotted line in Fig.\ref{fig:fig5}) is very close to the non-interacting Fermi surface. This is in strong contrast to other cases where the Luttinger surface created by interaction with the antiferromagnetic fluctuations intersect the Fermi surface to create small pockets in the nodal region\cite{YangRice06}.\\
\indent It is important to note that in order to obtain a good fit with the numerical result obtained by CDMFT, it is necessary here to take a value of $V^{\alpha}$ (in the superconducting phase) and $\Psi$ (in the pseudogap phase) significantly lower than the value of the pairing amplitude $\Delta$ in the superconducting phase. Indeed, an agreement between the experimental ARPES results and the fractionalized PDW scenario was obtained with the SC and CDW order being very close, a fact supported by Raman experiment in mercury-based cuprate\cite{Loret19}. This would indicate that only a part of the particle-particle pairs come from the fractionalization while the other part can be described by a standard superconducting condensate. This can also be compared to the two behaviours observed in Raman experiment for the nodal and antinodal part of the superconducting gap\cite{LeTacon06}. In fact, the nodal gap follows the critical temperature with doping and decreases away from optimal doping while the antinodal gap decreases linearly with doping, following the pseudogap temperature. This imbalance could be here a result of the finite size of the cluster (2x2 sites) used in the CDMFT calculation which is smaller than the wavelength of the charge modulation, thus hindering the formation of modulated orders.

\section{Discussion}
The results obtained from CDMFT calculations give valuable information on the energy dependence of the electronic Green's function in the doped Hubbard model. This is often limited to specific points in the Brillouin zone which depend on the cluster size. From this, it seems difficult to account for potential modulating orders with a wavelength larger than the cluster size. There is however a large number of experimental evidence\cite{Doiron-Leyraud07,Sebastian12,Chang12,Blanco-Canosa13,Blackburn13a,Ghiringhelli12,Gerber:2015gx,Chang:2016gz,Wu11,Wu:2015bt,Julien15,Loret19} for these orders to exist and compete in the pseudogap phase of cuprates superconductors. We argued in this paper that despite the numerical limitation, there are signatures of the charge order in the electronic self-energy calculated in CDMFT.\\\\
Much like in ARPES experiments, the observations are indirect consequences of the charge order and there is no Fermi surface reconstruction as the charge order remain short-ranged and fluctuating. In the case of ARPES in Bi2201, the main indicators for a modulating order were the back-bending of the band at the Brillouin zone's edge and the specific way the pseudogap was closing close to the nodal region (for more details see Ref.[\onlinecite{Lee14,He11}]). The limited momentum resolution of CDMFT does not allow for such distinction but the singular pole structure of the self-energy in the anti-nodal region could itself contain evidence for this modulating order. It is important to note the similarities between the original hidden-fermion from Ref.[\onlinecite{Sakai16}] and our proposition which contain the charge order. In particular both of them give the same result for $k\left(0,\pi\right)$. There are however distinctions to be made when going close to the nodal region where the two models differ. The main argument for the CDW scenario is that it allows for the double pole structure observed for $\omega>0$ (see Fig.\ref{fig:fig3}) which is observed numerically. This is to be taken with caution as both the numerical results and the idea of CDW order reaches their limit in the nodal region.\\\\
A more convincing argument may come from the electronic spectral function in the pseudogap phase. The original idea for the hidden-fermion is to describe the pseudogap by the absence of pairing. At the mean-field level, this would lead to a CDW long-range order with the associated Fermi surface reconstruction. The spectral function at the anti-nodal point does show the 2 poles observed by CDMFT with a vanishing weight at $\omega=0$ but there are still gapless states away from $k_x=0$. In contrast, the fractionalized PDW scenario is characterized by the equal superposition of SC and CDW fluctuations in the pseudogap. As such the spectral function has the same structure as in the superconducting state and is fully gapped at the Brillouin zone's edge, leaving Fermi arcs in the nodal region if we include a momentum dependence for the pseudogap order\cite{Grandadam20a,Grandadam20b}. Taking into account fluctuations in the SC and CDW orders leads to a drastic reduction in the spectral weight of specific poles, giving an effective spectral function with two apparent peaks. The same argument was given to explain the observation of a ``flat band'' in Bi2201 below $T_c$.\\\\
There are still many differences between the results from CDMFT and the experimental spectral function measured by ARPES. It is however interesting to find common ground to link these two essential techniques to investigate the pseudogap phase of cuprates. The capacity of CDMFT to explore the positive energies which are inaccessible to ARPES and the wide range of doping available to the experimental are complementary to elucidate the pseudogap phase of cuprates.

\section{ACKNOWLEDGMENTS}
We thank M. Civelli, A. Banerjee,  S. Sarkar and Y. Sidis for valuable discussions.   This work has received financial support from the ERC, under grant agreement  AdG-694651-CHAMPAGNE.

\bibliographystyle{apsrev4-1}
\bibliography{Cuprates}

\begin{thebibliography}{40}%
\makeatletter
\providecommand \@ifxundefined [1]{%
 \@ifx{#1\undefined}
}%
\providecommand \@ifnum [1]{%
 \ifnum #1\expandafter \@firstoftwo
 \else \expandafter \@secondoftwo
 \fi
}%
\providecommand \@ifx [1]{%
 \ifx #1\expandafter \@firstoftwo
 \else \expandafter \@secondoftwo
 \fi
}%
\providecommand \natexlab [1]{#1}%
\providecommand \enquote  [1]{``#1''}%
\providecommand \bibnamefont  [1]{#1}%
\providecommand \bibfnamefont [1]{#1}%
\providecommand \citenamefont [1]{#1}%
\providecommand \href@noop [0]{\@secondoftwo}%
\providecommand \href [0]{\begingroup \@sanitize@url \@href}%
\providecommand \@href[1]{\@@startlink{#1}\@@href}%
\providecommand \@@href[1]{\endgroup#1\@@endlink}%
\providecommand \@sanitize@url [0]{\catcode `\\12\catcode `\$12\catcode
  `\&12\catcode `\#12\catcode `\^12\catcode `\_12\catcode `\%12\relax}%
\providecommand \@@startlink[1]{}%
\providecommand \@@endlink[0]{}%
\providecommand \url  [0]{\begingroup\@sanitize@url \@url }%
\providecommand \@url [1]{\endgroup\@href {#1}{\urlprefix }}%
\providecommand \urlprefix  [0]{URL }%
\providecommand \Eprint [0]{\href }%
\providecommand \doibase [0]{http://dx.doi.org/}%
\providecommand \selectlanguage [0]{\@gobble}%
\providecommand \bibinfo  [0]{\@secondoftwo}%
\providecommand \bibfield  [0]{\@secondoftwo}%
\providecommand \translation [1]{[#1]}%
\providecommand \BibitemOpen [0]{}%
\providecommand \bibitemStop [0]{}%
\providecommand \bibitemNoStop [0]{.\EOS\space}%
\providecommand \EOS [0]{\spacefactor3000\relax}%
\providecommand \BibitemShut  [1]{\csname bibitem#1\endcsname}%
\let\auto@bib@innerbib\@empty
\bibitem [{\citenamefont {Sakai}\ \emph
  {et~al.}(2016{\natexlab{a}})\citenamefont {Sakai}, \citenamefont {Civelli},\
  and\ \citenamefont {Imada}}]{Sakai16}%
  \BibitemOpen
  \bibfield  {author} {\bibinfo {author} {\bibfnamefont {S.}~\bibnamefont
  {Sakai}}, \bibinfo {author} {\bibfnamefont {M.}~\bibnamefont {Civelli}}, \
  and\ \bibinfo {author} {\bibfnamefont {M.}~\bibnamefont {Imada}},\ }\href
  {\doibase 10.1103/PhysRevLett.116.057003} {\bibfield  {journal} {\bibinfo
  {journal} {Phys. Rev. Lett.}\ }\textbf {\bibinfo {volume} {116}},\ \bibinfo
  {pages} {057003} (\bibinfo {year} {2016}{\natexlab{a}})}\BibitemShut
  {NoStop}%
\bibitem [{\citenamefont {Alloul}\ \emph {et~al.}(1989)\citenamefont {Alloul},
  \citenamefont {Ohno},\ and\ \citenamefont {Mendels}}]{Alloul89}%
  \BibitemOpen
  \bibfield  {author} {\bibinfo {author} {\bibfnamefont {H.}~\bibnamefont
  {Alloul}}, \bibinfo {author} {\bibfnamefont {T.}~\bibnamefont {Ohno}}, \ and\
  \bibinfo {author} {\bibfnamefont {P.}~\bibnamefont {Mendels}},\ }\href
  {\doibase 10.1103/PhysRevLett.63.1700} {\bibfield  {journal} {\bibinfo
  {journal} {Phys. Rev. Lett.}\ }\textbf {\bibinfo {volume} {63}},\ \bibinfo
  {pages} {1700} (\bibinfo {year} {1989})}\BibitemShut {NoStop}%
\bibitem [{\citenamefont {Warren}\ \emph {et~al.}(1989)\citenamefont {Warren},
  \citenamefont {Walstedt}, \citenamefont {Brennert}, \citenamefont {Cava},
  \citenamefont {Tycko}, \citenamefont {Bell},\ and\ \citenamefont
  {Dabbagh}}]{Warren89}%
  \BibitemOpen
  \bibfield  {author} {\bibinfo {author} {\bibfnamefont {W.~W.}\ \bibnamefont
  {Warren}}, \bibinfo {author} {\bibfnamefont {R.~E.}\ \bibnamefont
  {Walstedt}}, \bibinfo {author} {\bibfnamefont {G.~F.}\ \bibnamefont
  {Brennert}}, \bibinfo {author} {\bibfnamefont {R.~J.}\ \bibnamefont {Cava}},
  \bibinfo {author} {\bibfnamefont {R.}~\bibnamefont {Tycko}}, \bibinfo
  {author} {\bibfnamefont {R.~F.}\ \bibnamefont {Bell}}, \ and\ \bibinfo
  {author} {\bibfnamefont {G.}~\bibnamefont {Dabbagh}},\ }\href {\doibase
  10.1103/PhysRevLett.62.1193} {\bibfield  {journal} {\bibinfo  {journal}
  {Phys. Rev. Lett.}\ }\textbf {\bibinfo {volume} {62}},\ \bibinfo {pages}
  {1193} (\bibinfo {year} {1989})}\BibitemShut {NoStop}%
\bibitem [{\citenamefont {Georges}\ \emph {et~al.}(1996)\citenamefont
  {Georges}, \citenamefont {Kotliar}, \citenamefont {Krauth},\ and\
  \citenamefont {Rozenberg}}]{Georges:1996tm}%
  \BibitemOpen
  \bibfield  {author} {\bibinfo {author} {\bibfnamefont {A.}~\bibnamefont
  {Georges}}, \bibinfo {author} {\bibfnamefont {G.}~\bibnamefont {Kotliar}},
  \bibinfo {author} {\bibfnamefont {W.}~\bibnamefont {Krauth}}, \ and\ \bibinfo
  {author} {\bibfnamefont {M.}~\bibnamefont {Rozenberg}},\ }\href {\doibase
  10.1103/RevModPhys.68.13} {\bibfield  {journal} {\bibinfo  {journal} {Reviews
  of Modern Physics}\ }\textbf {\bibinfo {volume} {68}},\ \bibinfo {pages} {13}
  (\bibinfo {year} {1996})}\BibitemShut {NoStop}%
\bibitem [{\citenamefont {Kotliar}\ \emph {et~al.}(2001)\citenamefont
  {Kotliar}, \citenamefont {Savrasov}, \citenamefont {Palsson},\ and\
  \citenamefont {Biroli}}]{Kotliar01}%
  \BibitemOpen
  \bibfield  {author} {\bibinfo {author} {\bibfnamefont {G.}~\bibnamefont
  {Kotliar}}, \bibinfo {author} {\bibfnamefont {S.~Y.}\ \bibnamefont
  {Savrasov}}, \bibinfo {author} {\bibfnamefont {G.}~\bibnamefont {Palsson}}, \
  and\ \bibinfo {author} {\bibfnamefont {G.}~\bibnamefont {Biroli}},\ }\href
  {\doibase 10.1103/PhysRevLett.87.186401} {\bibfield  {journal} {\bibinfo
  {journal} {Phys. Rev. Lett.}\ }\textbf {\bibinfo {volume} {87}},\ \bibinfo
  {pages} {186401} (\bibinfo {year} {2001})}\BibitemShut {NoStop}%
\bibitem [{\citenamefont {Maier}\ \emph {et~al.}(2005)\citenamefont {Maier},
  \citenamefont {Jarrell}, \citenamefont {Pruschke},\ and\ \citenamefont
  {Hettler}}]{Maier05}%
  \BibitemOpen
  \bibfield  {author} {\bibinfo {author} {\bibfnamefont {T.}~\bibnamefont
  {Maier}}, \bibinfo {author} {\bibfnamefont {M.}~\bibnamefont {Jarrell}},
  \bibinfo {author} {\bibfnamefont {T.}~\bibnamefont {Pruschke}}, \ and\
  \bibinfo {author} {\bibfnamefont {M.~H.}\ \bibnamefont {Hettler}},\ }\href
  {\doibase 10.1103/RevModPhys.77.1027} {\bibfield  {journal} {\bibinfo
  {journal} {Rev. Mod. Phys.}\ }\textbf {\bibinfo {volume} {77}},\ \bibinfo
  {pages} {1027} (\bibinfo {year} {2005})}\BibitemShut {NoStop}%
\bibitem [{\citenamefont {Gull}\ \emph {et~al.}(2010)\citenamefont {Gull},
  \citenamefont {Ferrero}, \citenamefont {Parcollet}, \citenamefont {Georges},\
  and\ \citenamefont {Millis}}]{Gull10}%
  \BibitemOpen
  \bibfield  {author} {\bibinfo {author} {\bibfnamefont {E.}~\bibnamefont
  {Gull}}, \bibinfo {author} {\bibfnamefont {M.}~\bibnamefont {Ferrero}},
  \bibinfo {author} {\bibfnamefont {O.}~\bibnamefont {Parcollet}}, \bibinfo
  {author} {\bibfnamefont {A.}~\bibnamefont {Georges}}, \ and\ \bibinfo
  {author} {\bibfnamefont {A.~J.}\ \bibnamefont {Millis}},\ }\href {\doibase
  10.1103/PhysRevB.82.155101} {\bibfield  {journal} {\bibinfo  {journal} {Phys.
  Rev. B}\ }\textbf {\bibinfo {volume} {82}},\ \bibinfo {pages} {155101}
  (\bibinfo {year} {2010})}\BibitemShut {NoStop}%
\bibitem [{\citenamefont {Sordi}\ \emph {et~al.}(2010)\citenamefont {Sordi},
  \citenamefont {Haule},\ and\ \citenamefont {Tremblay}}]{Sordi:2010iw}%
  \BibitemOpen
  \bibfield  {author} {\bibinfo {author} {\bibfnamefont {G.}~\bibnamefont
  {Sordi}}, \bibinfo {author} {\bibfnamefont {K.}~\bibnamefont {Haule}}, \ and\
  \bibinfo {author} {\bibfnamefont {A.}~\bibnamefont {Tremblay}},\ }\href
  {\doibase 10.1103/physrevlett.104.226402} {\bibfield  {journal} {\bibinfo
  {journal} {Phys. Rev. Lett.}\ }\textbf {\bibinfo {volume} {104}},\ \bibinfo
  {pages} {226402} (\bibinfo {year} {2010})}\BibitemShut {NoStop}%
\bibitem [{\citenamefont {Sordi}\ \emph {et~al.}(2012)\citenamefont {Sordi},
  \citenamefont {S{\'e}mon}, \citenamefont {Haule},\ and\ \citenamefont
  {Tremblay}}]{Sordi:2012bb}%
  \BibitemOpen
  \bibfield  {author} {\bibinfo {author} {\bibfnamefont {G.}~\bibnamefont
  {Sordi}}, \bibinfo {author} {\bibfnamefont {P.}~\bibnamefont {S{\'e}mon}},
  \bibinfo {author} {\bibfnamefont {K.}~\bibnamefont {Haule}}, \ and\ \bibinfo
  {author} {\bibfnamefont {A.}~\bibnamefont {Tremblay}},\ }\href {\doibase
  10.1103/physrevlett.108.216401} {\bibfield  {journal} {\bibinfo  {journal}
  {Phys. Rev. Lett.}\ }\textbf {\bibinfo {volume} {108}},\ \bibinfo {pages}
  {216401} (\bibinfo {year} {2012})}\BibitemShut {NoStop}%
\bibitem [{\citenamefont {Sakai}\ \emph
  {et~al.}(2016{\natexlab{b}})\citenamefont {Sakai}, \citenamefont {Civelli},\
  and\ \citenamefont {Imada}}]{Sakai:2016bo}%
  \BibitemOpen
  \bibfield  {author} {\bibinfo {author} {\bibfnamefont {S.}~\bibnamefont
  {Sakai}}, \bibinfo {author} {\bibfnamefont {M.}~\bibnamefont {Civelli}}, \
  and\ \bibinfo {author} {\bibfnamefont {M.}~\bibnamefont {Imada}},\
  }\href@noop {} {\bibfield  {journal} {\bibinfo  {journal} {Physical Review
  B}\ }\textbf {\bibinfo {volume} {94}},\ \bibinfo {pages} {115130} (\bibinfo
  {year} {2016}{\natexlab{b}})}\BibitemShut {NoStop}%
\bibitem [{\citenamefont {Sakai}\ \emph {et~al.}(2018)\citenamefont {Sakai},
  \citenamefont {Civelli},\ and\ \citenamefont {Imada}}]{Sakai:2018jq}%
  \BibitemOpen
  \bibfield  {author} {\bibinfo {author} {\bibfnamefont {S.}~\bibnamefont
  {Sakai}}, \bibinfo {author} {\bibfnamefont {M.}~\bibnamefont {Civelli}}, \
  and\ \bibinfo {author} {\bibfnamefont {M.}~\bibnamefont {Imada}},\
  }\href@noop {} {\bibfield  {journal} {\bibinfo  {journal} {Physical Review
  B}\ }\textbf {\bibinfo {volume} {98}},\ \bibinfo {pages} {195109} (\bibinfo
  {year} {2018})}\BibitemShut {NoStop}%
\bibitem [{\citenamefont {Scheurer}\ \emph {et~al.}(2018)\citenamefont
  {Scheurer}, \citenamefont {Chatterjee}, \citenamefont {Wu}, \citenamefont
  {Ferrero}, \citenamefont {Georges},\ and\ \citenamefont
  {Sachdev}}]{Scheurer18a}%
  \BibitemOpen
  \bibfield  {author} {\bibinfo {author} {\bibfnamefont {M.~S.}\ \bibnamefont
  {Scheurer}}, \bibinfo {author} {\bibfnamefont {S.}~\bibnamefont
  {Chatterjee}}, \bibinfo {author} {\bibfnamefont {W.}~\bibnamefont {Wu}},
  \bibinfo {author} {\bibfnamefont {M.}~\bibnamefont {Ferrero}}, \bibinfo
  {author} {\bibfnamefont {A.}~\bibnamefont {Georges}}, \ and\ \bibinfo
  {author} {\bibfnamefont {S.}~\bibnamefont {Sachdev}},\ }\href {\doibase
  10.1073/pnas.1720580115} {\bibfield  {journal} {\bibinfo  {journal}
  {Proceedings of the National Academy of Sciences}\ }\textbf {\bibinfo
  {volume} {115}},\ \bibinfo {pages} {E3665} (\bibinfo {year} {2018})},\
  \Eprint
  {http://arxiv.org/abs/https://www.pnas.org/content/115/16/E3665.full.pdf}
  {https://www.pnas.org/content/115/16/E3665.full.pdf} \BibitemShut {NoStop}%
\bibitem [{\citenamefont {Doiron-Leyraud}\ \emph {et~al.}(2007)\citenamefont
  {Doiron-Leyraud}, \citenamefont {Proust}, \citenamefont {LeBoeuf},
  \citenamefont {Levallois}, \citenamefont {Bonnemaison}, \citenamefont
  {Liang}, \citenamefont {Bonn}, \citenamefont {Hardy},\ and\ \citenamefont
  {Taillefer}}]{Doiron-Leyraud07}%
  \BibitemOpen
  \bibfield  {author} {\bibinfo {author} {\bibfnamefont {N.}~\bibnamefont
  {Doiron-Leyraud}}, \bibinfo {author} {\bibfnamefont {C.}~\bibnamefont
  {Proust}}, \bibinfo {author} {\bibfnamefont {D.}~\bibnamefont {LeBoeuf}},
  \bibinfo {author} {\bibfnamefont {J.}~\bibnamefont {Levallois}}, \bibinfo
  {author} {\bibfnamefont {J.-B.}\ \bibnamefont {Bonnemaison}}, \bibinfo
  {author} {\bibfnamefont {R.}~\bibnamefont {Liang}}, \bibinfo {author}
  {\bibfnamefont {D.~A.}\ \bibnamefont {Bonn}}, \bibinfo {author}
  {\bibfnamefont {W.~N.}\ \bibnamefont {Hardy}}, \ and\ \bibinfo {author}
  {\bibfnamefont {L.}~\bibnamefont {Taillefer}},\ }\href {\doibase
  10.1038/nature05872} {\bibfield  {journal} {\bibinfo  {journal} {Nature}\
  }\textbf {\bibinfo {volume} {447}},\ \bibinfo {pages} {565} (\bibinfo {year}
  {2007})}\BibitemShut {NoStop}%
\bibitem [{\citenamefont {Sebastian}\ \emph {et~al.}(2012)\citenamefont
  {Sebastian}, \citenamefont {Harrison}, \citenamefont {Liang}, \citenamefont
  {Bonn}, \citenamefont {Hardy}, \citenamefont {Mielke},\ and\ \citenamefont
  {Lonzarich}}]{Sebastian12}%
  \BibitemOpen
  \bibfield  {author} {\bibinfo {author} {\bibfnamefont {S.~E.}\ \bibnamefont
  {Sebastian}}, \bibinfo {author} {\bibfnamefont {N.}~\bibnamefont {Harrison}},
  \bibinfo {author} {\bibfnamefont {R.}~\bibnamefont {Liang}}, \bibinfo
  {author} {\bibfnamefont {D.~A.}\ \bibnamefont {Bonn}}, \bibinfo {author}
  {\bibfnamefont {W.~N.}\ \bibnamefont {Hardy}}, \bibinfo {author}
  {\bibfnamefont {C.~H.}\ \bibnamefont {Mielke}}, \ and\ \bibinfo {author}
  {\bibfnamefont {G.~G.}\ \bibnamefont {Lonzarich}},\ }\href {\doibase
  10.1103/PhysRevLett.108.196403} {\bibfield  {journal} {\bibinfo  {journal}
  {Phys. Rev. Lett.}\ }\textbf {\bibinfo {volume} {108}},\ \bibinfo {pages}
  {196403} (\bibinfo {year} {2012})}\BibitemShut {NoStop}%
\bibitem [{\citenamefont {Chang}\ \emph {et~al.}(2012)\citenamefont {Chang},
  \citenamefont {Blackburn}, \citenamefont {Holmes}, \citenamefont
  {Christensen}, \citenamefont {Larsen}, \citenamefont {Mesot}, \citenamefont
  {Liang}, \citenamefont {Bonn}, \citenamefont {Hardy}, \citenamefont
  {Watenphul}, \citenamefont {Zimmermann}, \citenamefont {Forgan},\ and\
  \citenamefont {Hayden}}]{Chang12}%
  \BibitemOpen
  \bibfield  {author} {\bibinfo {author} {\bibfnamefont {J.}~\bibnamefont
  {Chang}}, \bibinfo {author} {\bibfnamefont {E.}~\bibnamefont {Blackburn}},
  \bibinfo {author} {\bibfnamefont {A.~T.}\ \bibnamefont {Holmes}}, \bibinfo
  {author} {\bibfnamefont {N.~B.}\ \bibnamefont {Christensen}}, \bibinfo
  {author} {\bibfnamefont {J.}~\bibnamefont {Larsen}}, \bibinfo {author}
  {\bibfnamefont {J.}~\bibnamefont {Mesot}}, \bibinfo {author} {\bibfnamefont
  {R.}~\bibnamefont {Liang}}, \bibinfo {author} {\bibfnamefont {D.~A.}\
  \bibnamefont {Bonn}}, \bibinfo {author} {\bibfnamefont {W.~N.}\ \bibnamefont
  {Hardy}}, \bibinfo {author} {\bibfnamefont {A.}~\bibnamefont {Watenphul}},
  \bibinfo {author} {\bibfnamefont {M.~v.}\ \bibnamefont {Zimmermann}},
  \bibinfo {author} {\bibfnamefont {E.~M.}\ \bibnamefont {Forgan}}, \ and\
  \bibinfo {author} {\bibfnamefont {S.~M.}\ \bibnamefont {Hayden}},\ }\href
  {\doibase 10.1038/nphys2456} {\bibfield  {journal} {\bibinfo  {journal} {Nat.
  Phys.}\ }\textbf {\bibinfo {volume} {8}},\ \bibinfo {pages} {871} (\bibinfo
  {year} {2012})}\BibitemShut {NoStop}%
\bibitem [{\citenamefont {Blanco-Canosa}\ \emph {et~al.}(2013)\citenamefont
  {Blanco-Canosa}, \citenamefont {Frano}, \citenamefont {Loew}, \citenamefont
  {Lu}, \citenamefont {Porras}, \citenamefont {Ghiringhelli}, \citenamefont
  {Minola}, \citenamefont {Mazzoli}, \citenamefont {Braicovich}, \citenamefont
  {Schierle}, \citenamefont {Weschke}, \citenamefont {Le~Tacon},\ and\
  \citenamefont {Keimer}}]{Blanco-Canosa13}%
  \BibitemOpen
  \bibfield  {author} {\bibinfo {author} {\bibfnamefont {S.}~\bibnamefont
  {Blanco-Canosa}}, \bibinfo {author} {\bibfnamefont {A.}~\bibnamefont
  {Frano}}, \bibinfo {author} {\bibfnamefont {T.}~\bibnamefont {Loew}},
  \bibinfo {author} {\bibfnamefont {Y.}~\bibnamefont {Lu}}, \bibinfo {author}
  {\bibfnamefont {J.}~\bibnamefont {Porras}}, \bibinfo {author} {\bibfnamefont
  {G.}~\bibnamefont {Ghiringhelli}}, \bibinfo {author} {\bibfnamefont
  {M.}~\bibnamefont {Minola}}, \bibinfo {author} {\bibfnamefont
  {C.}~\bibnamefont {Mazzoli}}, \bibinfo {author} {\bibfnamefont
  {L.}~\bibnamefont {Braicovich}}, \bibinfo {author} {\bibfnamefont
  {E.}~\bibnamefont {Schierle}}, \bibinfo {author} {\bibfnamefont
  {E.}~\bibnamefont {Weschke}}, \bibinfo {author} {\bibfnamefont
  {M.}~\bibnamefont {Le~Tacon}}, \ and\ \bibinfo {author} {\bibfnamefont
  {B.}~\bibnamefont {Keimer}},\ }\href {\doibase
  10.1103/PhysRevLett.110.187001} {\bibfield  {journal} {\bibinfo  {journal}
  {Phys. Rev. Lett.}\ }\textbf {\bibinfo {volume} {110}},\ \bibinfo {pages}
  {187001} (\bibinfo {year} {2013})}\BibitemShut {NoStop}%
\bibitem [{\citenamefont {Blackburn}\ \emph {et~al.}(2013)\citenamefont
  {Blackburn}, \citenamefont {Chang}, \citenamefont {H\"ucker}, \citenamefont
  {Holmes}, \citenamefont {Christensen}, \citenamefont {Liang}, \citenamefont
  {Bonn}, \citenamefont {Hardy}, \citenamefont {R\"utt}, \citenamefont
  {Gutowski}, \citenamefont {Zimmermann}, \citenamefont {Forgan},\ and\
  \citenamefont {Hayden}}]{Blackburn13a}%
  \BibitemOpen
  \bibfield  {author} {\bibinfo {author} {\bibfnamefont {E.}~\bibnamefont
  {Blackburn}}, \bibinfo {author} {\bibfnamefont {J.}~\bibnamefont {Chang}},
  \bibinfo {author} {\bibfnamefont {M.}~\bibnamefont {H\"ucker}}, \bibinfo
  {author} {\bibfnamefont {A.~T.}\ \bibnamefont {Holmes}}, \bibinfo {author}
  {\bibfnamefont {N.~B.}\ \bibnamefont {Christensen}}, \bibinfo {author}
  {\bibfnamefont {R.}~\bibnamefont {Liang}}, \bibinfo {author} {\bibfnamefont
  {D.~A.}\ \bibnamefont {Bonn}}, \bibinfo {author} {\bibfnamefont {W.~N.}\
  \bibnamefont {Hardy}}, \bibinfo {author} {\bibfnamefont {U.}~\bibnamefont
  {R\"utt}}, \bibinfo {author} {\bibfnamefont {O.}~\bibnamefont {Gutowski}},
  \bibinfo {author} {\bibfnamefont {M.~v.}\ \bibnamefont {Zimmermann}},
  \bibinfo {author} {\bibfnamefont {E.~M.}\ \bibnamefont {Forgan}}, \ and\
  \bibinfo {author} {\bibfnamefont {S.~M.}\ \bibnamefont {Hayden}},\ }\href
  {\doibase 10.1103/PhysRevLett.110.137004} {\bibfield  {journal} {\bibinfo
  {journal} {Phys. Rev. Lett.}\ }\textbf {\bibinfo {volume} {110}},\ \bibinfo
  {pages} {137004} (\bibinfo {year} {2013})}\BibitemShut {NoStop}%
\bibitem [{\citenamefont {Ghiringhelli}\ \emph {et~al.}(2012)\citenamefont
  {Ghiringhelli}, \citenamefont {Le~Tacon}, \citenamefont {Minola},
  \citenamefont {Blanco-Canosa}, \citenamefont {Mazzoli}, \citenamefont
  {Brookes}, \citenamefont {De~Luca}, \citenamefont {Frano}, \citenamefont
  {Hawthorn}, \citenamefont {He}, \citenamefont {Loew}, \citenamefont {Sala},
  \citenamefont {Peets}, \citenamefont {Salluzzo}, \citenamefont {Schierle},
  \citenamefont {Sutarto}, \citenamefont {Sawatzky}, \citenamefont {Weschke},
  \citenamefont {Keimer},\ and\ \citenamefont {Braicovich}}]{Ghiringhelli12}%
  \BibitemOpen
  \bibfield  {author} {\bibinfo {author} {\bibfnamefont {G.}~\bibnamefont
  {Ghiringhelli}}, \bibinfo {author} {\bibfnamefont {M.}~\bibnamefont
  {Le~Tacon}}, \bibinfo {author} {\bibfnamefont {M.}~\bibnamefont {Minola}},
  \bibinfo {author} {\bibfnamefont {S.}~\bibnamefont {Blanco-Canosa}}, \bibinfo
  {author} {\bibfnamefont {C.}~\bibnamefont {Mazzoli}}, \bibinfo {author}
  {\bibfnamefont {N.~B.}\ \bibnamefont {Brookes}}, \bibinfo {author}
  {\bibfnamefont {G.~M.}\ \bibnamefont {De~Luca}}, \bibinfo {author}
  {\bibfnamefont {A.}~\bibnamefont {Frano}}, \bibinfo {author} {\bibfnamefont
  {D.~G.}\ \bibnamefont {Hawthorn}}, \bibinfo {author} {\bibfnamefont
  {F.}~\bibnamefont {He}}, \bibinfo {author} {\bibfnamefont {T.}~\bibnamefont
  {Loew}}, \bibinfo {author} {\bibfnamefont {M.~M.}\ \bibnamefont {Sala}},
  \bibinfo {author} {\bibfnamefont {D.~C.}\ \bibnamefont {Peets}}, \bibinfo
  {author} {\bibfnamefont {M.}~\bibnamefont {Salluzzo}}, \bibinfo {author}
  {\bibfnamefont {E.}~\bibnamefont {Schierle}}, \bibinfo {author}
  {\bibfnamefont {R.}~\bibnamefont {Sutarto}}, \bibinfo {author} {\bibfnamefont
  {G.~A.}\ \bibnamefont {Sawatzky}}, \bibinfo {author} {\bibfnamefont
  {E.}~\bibnamefont {Weschke}}, \bibinfo {author} {\bibfnamefont
  {B.}~\bibnamefont {Keimer}}, \ and\ \bibinfo {author} {\bibfnamefont
  {L.}~\bibnamefont {Braicovich}},\ }\href {\doibase 10.1126/science.1223532}
  {\bibfield  {journal} {\bibinfo  {journal} {Science}\ }\textbf {\bibinfo
  {volume} {337}},\ \bibinfo {pages} {821} (\bibinfo {year}
  {2012})}\BibitemShut {NoStop}%
\bibitem [{\citenamefont {Gerber}\ \emph {et~al.}(2015)\citenamefont {Gerber},
  \citenamefont {Jang}, \citenamefont {Nojiri}, \citenamefont {Matsuzawa},
  \citenamefont {Yasumura}, \citenamefont {Bonn}, \citenamefont {Liang},
  \citenamefont {Hardy}, \citenamefont {Islam}, \citenamefont {Mehta},
  \citenamefont {Song}, \citenamefont {Sikorski}, \citenamefont {Stefanescu},
  \citenamefont {Feng}, \citenamefont {Kivelson}, \citenamefont {Devereaux},
  \citenamefont {Shen}, \citenamefont {Kao}, \citenamefont {Lee}, \citenamefont
  {Zhu},\ and\ \citenamefont {Lee}}]{Gerber:2015gx}%
  \BibitemOpen
  \bibfield  {author} {\bibinfo {author} {\bibfnamefont {S.}~\bibnamefont
  {Gerber}}, \bibinfo {author} {\bibfnamefont {H.}~\bibnamefont {Jang}},
  \bibinfo {author} {\bibfnamefont {H.}~\bibnamefont {Nojiri}}, \bibinfo
  {author} {\bibfnamefont {S.}~\bibnamefont {Matsuzawa}}, \bibinfo {author}
  {\bibfnamefont {H.}~\bibnamefont {Yasumura}}, \bibinfo {author}
  {\bibfnamefont {D.~A.}\ \bibnamefont {Bonn}}, \bibinfo {author}
  {\bibfnamefont {R.}~\bibnamefont {Liang}}, \bibinfo {author} {\bibfnamefont
  {W.~N.}\ \bibnamefont {Hardy}}, \bibinfo {author} {\bibfnamefont
  {Z.}~\bibnamefont {Islam}}, \bibinfo {author} {\bibfnamefont
  {A.}~\bibnamefont {Mehta}}, \bibinfo {author} {\bibfnamefont
  {S.}~\bibnamefont {Song}}, \bibinfo {author} {\bibfnamefont {M.}~\bibnamefont
  {Sikorski}}, \bibinfo {author} {\bibfnamefont {D.}~\bibnamefont
  {Stefanescu}}, \bibinfo {author} {\bibfnamefont {Y.}~\bibnamefont {Feng}},
  \bibinfo {author} {\bibfnamefont {S.~A.}\ \bibnamefont {Kivelson}}, \bibinfo
  {author} {\bibfnamefont {T.~P.}\ \bibnamefont {Devereaux}}, \bibinfo {author}
  {\bibfnamefont {Z.-X.}\ \bibnamefont {Shen}}, \bibinfo {author}
  {\bibfnamefont {C.~C.}\ \bibnamefont {Kao}}, \bibinfo {author} {\bibfnamefont
  {W.~S.}\ \bibnamefont {Lee}}, \bibinfo {author} {\bibfnamefont
  {D.}~\bibnamefont {Zhu}}, \ and\ \bibinfo {author} {\bibfnamefont {J.~S.}\
  \bibnamefont {Lee}},\ }\href {\doibase 10.1126/science.aac6257} {\bibfield
  {journal} {\bibinfo  {journal} {Science}\ }\textbf {\bibinfo {volume}
  {350}},\ \bibinfo {pages} {949} (\bibinfo {year} {2015})}\BibitemShut
  {NoStop}%
\bibitem [{\citenamefont {Chang}\ \emph
  {et~al.}(2016{\natexlab{a}})\citenamefont {Chang}, \citenamefont {Blackburn},
  \citenamefont {Ivashko}, \citenamefont {Holmes}, \citenamefont {Christensen},
  \citenamefont {Huecker}, \citenamefont {Liang}, \citenamefont {Bonn},
  \citenamefont {Hardy}, \citenamefont {Ruett}, \citenamefont {Zimmermann},
  \citenamefont {Forgan},\ and\ \citenamefont {Hayden}}]{Chang:2016gz}%
  \BibitemOpen
  \bibfield  {author} {\bibinfo {author} {\bibfnamefont {J.}~\bibnamefont
  {Chang}}, \bibinfo {author} {\bibfnamefont {E.}~\bibnamefont {Blackburn}},
  \bibinfo {author} {\bibfnamefont {O.}~\bibnamefont {Ivashko}}, \bibinfo
  {author} {\bibfnamefont {A.~T.}\ \bibnamefont {Holmes}}, \bibinfo {author}
  {\bibfnamefont {N.~B.}\ \bibnamefont {Christensen}}, \bibinfo {author}
  {\bibfnamefont {M.}~\bibnamefont {Huecker}}, \bibinfo {author} {\bibfnamefont
  {R.}~\bibnamefont {Liang}}, \bibinfo {author} {\bibfnamefont {D.~A.}\
  \bibnamefont {Bonn}}, \bibinfo {author} {\bibfnamefont {W.~N.}\ \bibnamefont
  {Hardy}}, \bibinfo {author} {\bibfnamefont {U.}~\bibnamefont {Ruett}},
  \bibinfo {author} {\bibfnamefont {M.~V.}\ \bibnamefont {Zimmermann}},
  \bibinfo {author} {\bibfnamefont {E.~M.}\ \bibnamefont {Forgan}}, \ and\
  \bibinfo {author} {\bibfnamefont {S.~M.}\ \bibnamefont {Hayden}},\ }\href
  {\doibase 10.1038/ncomms11494} {\bibfield  {journal} {\bibinfo  {journal}
  {Nat. Commun.}\ }\textbf {\bibinfo {volume} {7}} (\bibinfo {year}
  {2016}{\natexlab{a}}),\ 10.1038/ncomms11494}\BibitemShut {NoStop}%
\bibitem [{\citenamefont {Wu}\ \emph {et~al.}(2011)\citenamefont {Wu},
  \citenamefont {Mayaffre}, \citenamefont {Kr{\"a}mer}, \citenamefont
  {Horvatic}, \citenamefont {Berthier}, \citenamefont {Hardy}, \citenamefont
  {Liang}, \citenamefont {Bonn},\ and\ \citenamefont {Julien}}]{Wu11}%
  \BibitemOpen
  \bibfield  {author} {\bibinfo {author} {\bibfnamefont {T.}~\bibnamefont
  {Wu}}, \bibinfo {author} {\bibfnamefont {H.}~\bibnamefont {Mayaffre}},
  \bibinfo {author} {\bibfnamefont {S.}~\bibnamefont {Kr{\"a}mer}}, \bibinfo
  {author} {\bibfnamefont {M.}~\bibnamefont {Horvatic}}, \bibinfo {author}
  {\bibfnamefont {C.}~\bibnamefont {Berthier}}, \bibinfo {author}
  {\bibfnamefont {W.~N.}\ \bibnamefont {Hardy}}, \bibinfo {author}
  {\bibfnamefont {R.}~\bibnamefont {Liang}}, \bibinfo {author} {\bibfnamefont
  {D.~A.}\ \bibnamefont {Bonn}}, \ and\ \bibinfo {author} {\bibfnamefont
  {M.-H.}\ \bibnamefont {Julien}},\ }\href {\doibase 10.1038/nature10345}
  {\bibfield  {journal} {\bibinfo  {journal} {Nature}\ }\textbf {\bibinfo
  {volume} {477}},\ \bibinfo {pages} {191} (\bibinfo {year}
  {2011})}\BibitemShut {NoStop}%
\bibitem [{\citenamefont {{Wu}}\ \emph {et~al.}(2015)\citenamefont {{Wu}},
  \citenamefont {{Mayaffre}}, \citenamefont {{Kr{\"a}mer}}, \citenamefont
  {{Horvati{\'c}}}, \citenamefont {{Berthier}}, \citenamefont {{Hardy}},
  \citenamefont {{Liang}}, \citenamefont {{Bonn}},\ and\ \citenamefont
  {{Julien}}}]{Wu:2015bt}%
  \BibitemOpen
  \bibfield  {author} {\bibinfo {author} {\bibfnamefont {T.}~\bibnamefont
  {{Wu}}}, \bibinfo {author} {\bibfnamefont {H.}~\bibnamefont {{Mayaffre}}},
  \bibinfo {author} {\bibfnamefont {S.}~\bibnamefont {{Kr{\"a}mer}}}, \bibinfo
  {author} {\bibfnamefont {M.}~\bibnamefont {{Horvati{\'c}}}}, \bibinfo
  {author} {\bibfnamefont {C.}~\bibnamefont {{Berthier}}}, \bibinfo {author}
  {\bibfnamefont {W.~N.}\ \bibnamefont {{Hardy}}}, \bibinfo {author}
  {\bibfnamefont {R.}~\bibnamefont {{Liang}}}, \bibinfo {author} {\bibfnamefont
  {D.~A.}\ \bibnamefont {{Bonn}}}, \ and\ \bibinfo {author} {\bibfnamefont
  {M.-H.}\ \bibnamefont {{Julien}}},\ }\href {\doibase 10.1038/ncomms7438}
  {\bibfield  {journal} {\bibinfo  {journal} {Nature Communications}\ }\textbf
  {\bibinfo {volume} {6}},\ \bibinfo {eid} {6438} (\bibinfo {year}
  {2015})}\BibitemShut {NoStop}%
\bibitem [{\citenamefont {Julien}(2015)}]{Julien15}%
  \BibitemOpen
  \bibfield  {author} {\bibinfo {author} {\bibfnamefont {M.-H.}\ \bibnamefont
  {Julien}},\ }\href {\doibase 10.1126/science.aad3279} {\bibfield  {journal}
  {\bibinfo  {journal} {Science}\ }\textbf {\bibinfo {volume} {350}},\ \bibinfo
  {pages} {914} (\bibinfo {year} {2015})}\BibitemShut {NoStop}%
\bibitem [{\citenamefont {Loret}\ \emph {et~al.}(2019)\citenamefont {Loret},
  \citenamefont {Auvray}, \citenamefont {Gallais}, \citenamefont {Cazayous},
  \citenamefont {Forget}, \citenamefont {Colson}, \citenamefont {Julien},
  \citenamefont {Paul}, \citenamefont {Civelli},\ and\ \citenamefont
  {Sacuto}}]{Loret19}%
  \BibitemOpen
  \bibfield  {author} {\bibinfo {author} {\bibfnamefont {B.}~\bibnamefont
  {Loret}}, \bibinfo {author} {\bibfnamefont {N.}~\bibnamefont {Auvray}},
  \bibinfo {author} {\bibfnamefont {Y.}~\bibnamefont {Gallais}}, \bibinfo
  {author} {\bibfnamefont {M.}~\bibnamefont {Cazayous}}, \bibinfo {author}
  {\bibfnamefont {A.}~\bibnamefont {Forget}}, \bibinfo {author} {\bibfnamefont
  {D.}~\bibnamefont {Colson}}, \bibinfo {author} {\bibfnamefont {M.-H.}\
  \bibnamefont {Julien}}, \bibinfo {author} {\bibfnamefont {I.}~\bibnamefont
  {Paul}}, \bibinfo {author} {\bibfnamefont {M.}~\bibnamefont {Civelli}}, \
  and\ \bibinfo {author} {\bibfnamefont {A.}~\bibnamefont {Sacuto}},\ }\href
  {https://www.nature.com/articles/s41567-019-0509-5} {\bibfield  {journal}
  {\bibinfo  {journal} {Nature Physics}\ ,\ \bibinfo {pages} {1}} (\bibinfo
  {year} {2019})}\BibitemShut {NoStop}%
\bibitem [{\citenamefont {LeBoeuf}\ \emph {et~al.}(2013)\citenamefont
  {LeBoeuf}, \citenamefont {Kramer}, \citenamefont {Hardy}, \citenamefont
  {Liang}, \citenamefont {Bonn},\ and\ \citenamefont {Proust}}]{LeBoeuf13}%
  \BibitemOpen
  \bibfield  {author} {\bibinfo {author} {\bibfnamefont {D.}~\bibnamefont
  {LeBoeuf}}, \bibinfo {author} {\bibfnamefont {S.}~\bibnamefont {Kramer}},
  \bibinfo {author} {\bibfnamefont {W.~N.}\ \bibnamefont {Hardy}}, \bibinfo
  {author} {\bibfnamefont {R.}~\bibnamefont {Liang}}, \bibinfo {author}
  {\bibfnamefont {D.~A.}\ \bibnamefont {Bonn}}, \ and\ \bibinfo {author}
  {\bibfnamefont {C.}~\bibnamefont {Proust}},\ }\href
  {http://dx.doi.org/10.1038/nphys2502} {\bibfield  {journal} {\bibinfo
  {journal} {Nat. Phys.}\ }\textbf {\bibinfo {volume} {9}},\ \bibinfo {pages}
  {79} (\bibinfo {year} {2013})}\BibitemShut {NoStop}%
\bibitem [{\citenamefont {Chakraborty}\ \emph {et~al.}(2019)\citenamefont
  {Chakraborty}, \citenamefont {Grandadam}, \citenamefont {Hamidian},
  \citenamefont {Davis}, \citenamefont {Sidis},\ and\ \citenamefont
  {P\'epin}}]{Chakraborty19}%
  \BibitemOpen
  \bibfield  {author} {\bibinfo {author} {\bibfnamefont {D.}~\bibnamefont
  {Chakraborty}}, \bibinfo {author} {\bibfnamefont {M.}~\bibnamefont
  {Grandadam}}, \bibinfo {author} {\bibfnamefont {M.~H.}\ \bibnamefont
  {Hamidian}}, \bibinfo {author} {\bibfnamefont {J.~C.~S.}\ \bibnamefont
  {Davis}}, \bibinfo {author} {\bibfnamefont {Y.}~\bibnamefont {Sidis}}, \ and\
  \bibinfo {author} {\bibfnamefont {C.}~\bibnamefont {P\'epin}},\ }\href
  {\doibase 10.1103/PhysRevB.100.224511} {\bibfield  {journal} {\bibinfo
  {journal} {Phys. Rev. B}\ }\textbf {\bibinfo {volume} {100}},\ \bibinfo
  {pages} {224511} (\bibinfo {year} {2019})}\BibitemShut {NoStop}%
\bibitem [{\citenamefont {Grandadam}\ \emph
  {et~al.}(2020{\natexlab{a}})\citenamefont {Grandadam}, \citenamefont
  {Chakraborty},\ and\ \citenamefont {Pépin}}]{Grandadam20a}%
  \BibitemOpen
  \bibfield  {author} {\bibinfo {author} {\bibfnamefont {M.}~\bibnamefont
  {Grandadam}}, \bibinfo {author} {\bibfnamefont {D.}~\bibnamefont
  {Chakraborty}}, \ and\ \bibinfo {author} {\bibfnamefont {C.}~\bibnamefont
  {Pépin}},\ }\href {\doibase 10.1007/s10948-019-05380-6} {\bibfield
  {journal} {\bibinfo  {journal} {J. Supercond. Nov. Magn.}\ }\textbf {\bibinfo
  {volume} {33}},\ \bibinfo {pages} {2361} (\bibinfo {year}
  {2020}{\natexlab{a}})}\BibitemShut {NoStop}%
\bibitem [{\citenamefont {Grandadam}\ \emph
  {et~al.}(2020{\natexlab{b}})\citenamefont {Grandadam}, \citenamefont
  {Chakraborty}, \citenamefont {Montiel},\ and\ \citenamefont
  {P\'epin}}]{Grandadam20b}%
  \BibitemOpen
  \bibfield  {author} {\bibinfo {author} {\bibfnamefont {M.}~\bibnamefont
  {Grandadam}}, \bibinfo {author} {\bibfnamefont {D.}~\bibnamefont
  {Chakraborty}}, \bibinfo {author} {\bibfnamefont {X.}~\bibnamefont
  {Montiel}}, \ and\ \bibinfo {author} {\bibfnamefont {C.}~\bibnamefont
  {P\'epin}},\ }\href {\doibase 10.1103/PhysRevB.102.121104} {\bibfield
  {journal} {\bibinfo  {journal} {Phys. Rev. B}\ }\textbf {\bibinfo {volume}
  {102}},\ \bibinfo {pages} {121104} (\bibinfo {year}
  {2020}{\natexlab{b}})}\BibitemShut {NoStop}%
\bibitem [{\citenamefont {He}\ \emph {et~al.}(2011)\citenamefont {He},
  \citenamefont {Hashimoto}, \citenamefont {Karapetyan}, \citenamefont
  {Koralek}, \citenamefont {Hinton}, \citenamefont {Testaud}, \citenamefont
  {Nathan}, \citenamefont {Yoshida}, \citenamefont {Yao}, \citenamefont
  {Tanaka}, \citenamefont {Meevasana}, \citenamefont {Moore}, \citenamefont
  {Lu}, \citenamefont {Mo}, \citenamefont {Ishikado}, \citenamefont {Eisaki},
  \citenamefont {Hussain}, \citenamefont {Devereaux}, \citenamefont {Kivelson},
  \citenamefont {Orenstein}, \citenamefont {Kapitulnik},\ and\ \citenamefont
  {Shen}}]{He11}%
  \BibitemOpen
  \bibfield  {author} {\bibinfo {author} {\bibfnamefont {R.-H.}\ \bibnamefont
  {He}}, \bibinfo {author} {\bibfnamefont {M.}~\bibnamefont {Hashimoto}},
  \bibinfo {author} {\bibfnamefont {H.}~\bibnamefont {Karapetyan}}, \bibinfo
  {author} {\bibfnamefont {J.~D.}\ \bibnamefont {Koralek}}, \bibinfo {author}
  {\bibfnamefont {J.~P.}\ \bibnamefont {Hinton}}, \bibinfo {author}
  {\bibfnamefont {J.~P.}\ \bibnamefont {Testaud}}, \bibinfo {author}
  {\bibfnamefont {V.}~\bibnamefont {Nathan}}, \bibinfo {author} {\bibfnamefont
  {Y.}~\bibnamefont {Yoshida}}, \bibinfo {author} {\bibfnamefont
  {H.}~\bibnamefont {Yao}}, \bibinfo {author} {\bibfnamefont {K.}~\bibnamefont
  {Tanaka}}, \bibinfo {author} {\bibfnamefont {W.}~\bibnamefont {Meevasana}},
  \bibinfo {author} {\bibfnamefont {R.~G.}\ \bibnamefont {Moore}}, \bibinfo
  {author} {\bibfnamefont {D.~H.}\ \bibnamefont {Lu}}, \bibinfo {author}
  {\bibfnamefont {S.-K.}\ \bibnamefont {Mo}}, \bibinfo {author} {\bibfnamefont
  {M.}~\bibnamefont {Ishikado}}, \bibinfo {author} {\bibfnamefont
  {H.}~\bibnamefont {Eisaki}}, \bibinfo {author} {\bibfnamefont
  {Z.}~\bibnamefont {Hussain}}, \bibinfo {author} {\bibfnamefont {T.~P.}\
  \bibnamefont {Devereaux}}, \bibinfo {author} {\bibfnamefont {S.~A.}\
  \bibnamefont {Kivelson}}, \bibinfo {author} {\bibfnamefont {J.}~\bibnamefont
  {Orenstein}}, \bibinfo {author} {\bibfnamefont {A.}~\bibnamefont
  {Kapitulnik}}, \ and\ \bibinfo {author} {\bibfnamefont {Z.-X.}\ \bibnamefont
  {Shen}},\ }\href {\doibase 10.1126/science.1198415} {\bibfield  {journal}
  {\bibinfo  {journal} {Science}\ }\textbf {\bibinfo {volume} {331}},\ \bibinfo
  {pages} {1579} (\bibinfo {year} {2011})}\BibitemShut {NoStop}%
\bibitem [{\citenamefont {Haule}\ and\ \citenamefont
  {Kotliar}(2007)}]{Haule07}%
  \BibitemOpen
  \bibfield  {author} {\bibinfo {author} {\bibfnamefont {K.}~\bibnamefont
  {Haule}}\ and\ \bibinfo {author} {\bibfnamefont {G.}~\bibnamefont
  {Kotliar}},\ }\href {\doibase 10.1103/PhysRevB.76.104509} {\bibfield
  {journal} {\bibinfo  {journal} {Phys. Rev. B}\ }\textbf {\bibinfo {volume}
  {76}},\ \bibinfo {pages} {104509} (\bibinfo {year} {2007})}\BibitemShut
  {NoStop}%
\bibitem [{\citenamefont {Gull}\ \emph {et~al.}(2013)\citenamefont {Gull},
  \citenamefont {Parcollet},\ and\ \citenamefont {Millis}}]{Gull:2013hh}%
  \BibitemOpen
  \bibfield  {author} {\bibinfo {author} {\bibfnamefont {E.}~\bibnamefont
  {Gull}}, \bibinfo {author} {\bibfnamefont {O.}~\bibnamefont {Parcollet}}, \
  and\ \bibinfo {author} {\bibfnamefont {A.~J.}\ \bibnamefont {Millis}},\
  }\href {\doibase 10.1103/PhysRevLett.110.216405} {\bibfield  {journal}
  {\bibinfo  {journal} {Phys. Rev. Lett.}\ }\textbf {\bibinfo {volume} {110}},\
  \bibinfo {pages} {216405} (\bibinfo {year} {2013})}\BibitemShut {NoStop}%
\bibitem [{\citenamefont {{Campi}}\ \emph {et~al.}(2015)\citenamefont
  {{Campi}}, \citenamefont {{Bianconi}}, \citenamefont {{Poccia}},
  \citenamefont {{Bianconi}}, \citenamefont {{Barba}}, \citenamefont
  {{Arrighetti}}, \citenamefont {{Innocenti}}, \citenamefont {{Karpinski}},
  \citenamefont {{Zhigadlo}}, \citenamefont {{Kazakov}}, \citenamefont
  {{Burghammer}}, \citenamefont {{Zimmermann}}, \citenamefont {{Sprung}},\ and\
  \citenamefont {{Ricci}}}]{Campi15}%
  \BibitemOpen
  \bibfield  {author} {\bibinfo {author} {\bibfnamefont {G.}~\bibnamefont
  {{Campi}}}, \bibinfo {author} {\bibfnamefont {A.}~\bibnamefont {{Bianconi}}},
  \bibinfo {author} {\bibfnamefont {N.}~\bibnamefont {{Poccia}}}, \bibinfo
  {author} {\bibfnamefont {G.}~\bibnamefont {{Bianconi}}}, \bibinfo {author}
  {\bibfnamefont {L.}~\bibnamefont {{Barba}}}, \bibinfo {author} {\bibfnamefont
  {G.}~\bibnamefont {{Arrighetti}}}, \bibinfo {author} {\bibfnamefont
  {D.}~\bibnamefont {{Innocenti}}}, \bibinfo {author} {\bibfnamefont
  {J.}~\bibnamefont {{Karpinski}}}, \bibinfo {author} {\bibfnamefont {N.~D.}\
  \bibnamefont {{Zhigadlo}}}, \bibinfo {author} {\bibfnamefont {S.~M.}\
  \bibnamefont {{Kazakov}}}, \bibinfo {author} {\bibfnamefont {M.}~\bibnamefont
  {{Burghammer}}}, \bibinfo {author} {\bibfnamefont {M.~V.}\ \bibnamefont
  {{Zimmermann}}}, \bibinfo {author} {\bibfnamefont {M.}~\bibnamefont
  {{Sprung}}}, \ and\ \bibinfo {author} {\bibfnamefont {A.}~\bibnamefont
  {{Ricci}}},\ }\href {\doibase 10.1038/nature14987} {\bibfield  {journal}
  {\bibinfo  {journal} {\nat}\ }\textbf {\bibinfo {volume} {525}},\ \bibinfo
  {pages} {359} (\bibinfo {year} {2015})},\ \Eprint
  {http://arxiv.org/abs/1509.05002} {arXiv:1509.05002 [cond-mat.supr-con]}
  \BibitemShut {NoStop}%
\bibitem [{\citenamefont {Chang}\ \emph
  {et~al.}(2016{\natexlab{b}})\citenamefont {Chang}, \citenamefont {Blackburn},
  \citenamefont {Ivashko}, \citenamefont {Holmes}, \citenamefont {Christensen},
  \citenamefont {Hucker}, \citenamefont {Liang}, \citenamefont {Bonn},
  \citenamefont {Hardy}, \citenamefont {Rutt}, \citenamefont {Zimmermann},
  \citenamefont {Forgan},\ and\ \citenamefont {M.}}]{Chang16}%
  \BibitemOpen
  \bibfield  {author} {\bibinfo {author} {\bibfnamefont {J.}~\bibnamefont
  {Chang}}, \bibinfo {author} {\bibfnamefont {E.}~\bibnamefont {Blackburn}},
  \bibinfo {author} {\bibfnamefont {O.}~\bibnamefont {Ivashko}}, \bibinfo
  {author} {\bibfnamefont {A.~T.}\ \bibnamefont {Holmes}}, \bibinfo {author}
  {\bibfnamefont {N.~B.}\ \bibnamefont {Christensen}}, \bibinfo {author}
  {\bibfnamefont {M.}~\bibnamefont {Hucker}}, \bibinfo {author} {\bibfnamefont
  {R.}~\bibnamefont {Liang}}, \bibinfo {author} {\bibfnamefont {D.~A.}\
  \bibnamefont {Bonn}}, \bibinfo {author} {\bibfnamefont {W.~N.}\ \bibnamefont
  {Hardy}}, \bibinfo {author} {\bibfnamefont {U.}~\bibnamefont {Rutt}},
  \bibinfo {author} {\bibfnamefont {M.~v.}\ \bibnamefont {Zimmermann}},
  \bibinfo {author} {\bibfnamefont {E.~M.}\ \bibnamefont {Forgan}}, \ and\
  \bibinfo {author} {\bibfnamefont {H.~S.}\ \bibnamefont {M.}},\ }\href
  {http://http://www.nature.com/articles/ncomms11494} {\bibfield  {journal}
  {\bibinfo  {journal} {Nat. Commun.}\ }\textbf {\bibinfo {volume} {7}},\
  \bibinfo {pages} {11494} (\bibinfo {year} {2016}{\natexlab{b}})}\BibitemShut
  {NoStop}%
\bibitem [{\citenamefont {Wu}\ \emph {et~al.}(2013)\citenamefont {Wu},
  \citenamefont {Mayaffre}, \citenamefont {Kr{\"a}mer}, \citenamefont
  {Horvati{\'c}}, \citenamefont {Berthier}, \citenamefont {Kuhns},
  \citenamefont {Reyes}, \citenamefont {Liang}, \citenamefont {Hardy},
  \citenamefont {Bonn},\ and\ \citenamefont {Julien}}]{Wu13a}%
  \BibitemOpen
  \bibfield  {author} {\bibinfo {author} {\bibfnamefont {T.}~\bibnamefont
  {Wu}}, \bibinfo {author} {\bibfnamefont {H.}~\bibnamefont {Mayaffre}},
  \bibinfo {author} {\bibfnamefont {S.}~\bibnamefont {Kr{\"a}mer}}, \bibinfo
  {author} {\bibfnamefont {M.}~\bibnamefont {Horvati{\'c}}}, \bibinfo {author}
  {\bibfnamefont {C.}~\bibnamefont {Berthier}}, \bibinfo {author}
  {\bibfnamefont {P.~L.}\ \bibnamefont {Kuhns}}, \bibinfo {author}
  {\bibfnamefont {A.~P.}\ \bibnamefont {Reyes}}, \bibinfo {author}
  {\bibfnamefont {R.}~\bibnamefont {Liang}}, \bibinfo {author} {\bibfnamefont
  {W.~N.}\ \bibnamefont {Hardy}}, \bibinfo {author} {\bibfnamefont {D.~A.}\
  \bibnamefont {Bonn}}, \ and\ \bibinfo {author} {\bibfnamefont {M.-H.}\
  \bibnamefont {Julien}},\ }\href {\doibase 10.1038/ncomms3113} {\bibfield
  {journal} {\bibinfo  {journal} {Nat. Commun.}\ }\textbf {\bibinfo {volume}
  {4}},\ \bibinfo {pages} {2113} (\bibinfo {year} {2013})}\BibitemShut
  {NoStop}%
\bibitem [{\citenamefont {Sarkar}\ \emph {et~al.}(2020)\citenamefont {Sarkar},
  \citenamefont {Grandadam},\ and\ \citenamefont {Pépin}}]{Sarkar20}%
  \BibitemOpen
  \bibfield  {author} {\bibinfo {author} {\bibfnamefont {S.}~\bibnamefont
  {Sarkar}}, \bibinfo {author} {\bibfnamefont {M.}~\bibnamefont {Grandadam}}, \
  and\ \bibinfo {author} {\bibfnamefont {C.}~\bibnamefont {Pépin}},\
  }\href@noop {} {\  (\bibinfo {year} {2020})},\ \Eprint
  {http://arxiv.org/abs/2009.02975} {arXiv:2009.02975 [cond-mat.supr-con]}
  \BibitemShut {NoStop}%
\bibitem [{\citenamefont {Baskaran}\ and\ \citenamefont
  {Anderson}(1988)}]{Baskaran88}%
  \BibitemOpen
  \bibfield  {author} {\bibinfo {author} {\bibfnamefont {G.}~\bibnamefont
  {Baskaran}}\ and\ \bibinfo {author} {\bibfnamefont {P.~W.}\ \bibnamefont
  {Anderson}},\ }\href {\doibase 10.1103/PhysRevB.37.580} {\bibfield  {journal}
  {\bibinfo  {journal} {Phys. Rev. B}\ }\textbf {\bibinfo {volume} {37}},\
  \bibinfo {pages} {580} (\bibinfo {year} {1988})}\BibitemShut {NoStop}%
\bibitem [{\citenamefont {Lee}\ and\ \citenamefont {Nagaosa}(1992)}]{Lee92}%
  \BibitemOpen
  \bibfield  {author} {\bibinfo {author} {\bibfnamefont {P.~A.}\ \bibnamefont
  {Lee}}\ and\ \bibinfo {author} {\bibfnamefont {N.}~\bibnamefont {Nagaosa}},\
  }\href {\doibase 10.1103/PhysRevB.46.5621} {\bibfield  {journal} {\bibinfo
  {journal} {Phys. Rev. B}\ }\textbf {\bibinfo {volume} {46}},\ \bibinfo
  {pages} {5621} (\bibinfo {year} {1992})}\BibitemShut {NoStop}%
\bibitem [{\citenamefont {Yang}\ \emph {et~al.}(2006)\citenamefont {Yang},
  \citenamefont {Rice},\ and\ \citenamefont {Zhang}}]{YangRice06}%
  \BibitemOpen
  \bibfield  {author} {\bibinfo {author} {\bibfnamefont {K.-Y.}\ \bibnamefont
  {Yang}}, \bibinfo {author} {\bibfnamefont {T.~M.}\ \bibnamefont {Rice}}, \
  and\ \bibinfo {author} {\bibfnamefont {F.-C.}\ \bibnamefont {Zhang}},\ }\href
  {\doibase 10.1103/PhysRevB.73.174501} {\bibfield  {journal} {\bibinfo
  {journal} {Phys. Rev. B}\ }\textbf {\bibinfo {volume} {73}},\ \bibinfo
  {pages} {174501} (\bibinfo {year} {2006})}\BibitemShut {NoStop}%
\bibitem [{\citenamefont {Le~Tacon}\ \emph {et~al.}(2006)\citenamefont
  {Le~Tacon}, \citenamefont {Sacuto}, \citenamefont {Georges}, \citenamefont
  {Kotliar}, \citenamefont {Gallais}, \citenamefont {Colson},\ and\
  \citenamefont {Forget}}]{LeTacon06}%
  \BibitemOpen
  \bibfield  {author} {\bibinfo {author} {\bibfnamefont {M.}~\bibnamefont
  {Le~Tacon}}, \bibinfo {author} {\bibfnamefont {A.}~\bibnamefont {Sacuto}},
  \bibinfo {author} {\bibfnamefont {A.}~\bibnamefont {Georges}}, \bibinfo
  {author} {\bibfnamefont {G.}~\bibnamefont {Kotliar}}, \bibinfo {author}
  {\bibfnamefont {Y.}~\bibnamefont {Gallais}}, \bibinfo {author} {\bibfnamefont
  {D.}~\bibnamefont {Colson}}, \ and\ \bibinfo {author} {\bibfnamefont
  {A.}~\bibnamefont {Forget}},\ }\href {\doibase 10.1038/nphys362} {\bibfield
  {journal} {\bibinfo  {journal} {Nat. Phys.}\ }\textbf {\bibinfo {volume}
  {2}},\ \bibinfo {pages} {537} (\bibinfo {year} {2006})}\BibitemShut {NoStop}%
\bibitem [{\citenamefont {Lee}(2014)}]{Lee14}%
  \BibitemOpen
  \bibfield  {author} {\bibinfo {author} {\bibfnamefont {P.~A.}\ \bibnamefont
  {Lee}},\ }\href {\doibase 10.1103/PhysRevX.4.031017} {\bibfield  {journal}
  {\bibinfo  {journal} {Phys. Rev. X}\ }\textbf {\bibinfo {volume} {4}},\
  \bibinfo {pages} {031017} (\bibinfo {year} {2014})}\BibitemShut {NoStop}%
\end{thebibliography}%

\end{document}